\documentclass[11pt,twoside]{article}
\usepackage{./asp2014}

\aspSuppressVolSlug
\resetcounters

\begin{document}

\title{Potential for Solar System Science with the ngVLA}
\author{Imke de Pater,$^1$ Bryan Butler,$^2$ R. J. Sault, $^3$, Arielle Moullet, $^4$, Chris Moeckel, $^5$, Joshua Tollefson, $^6$, Katherine de Kleer, $^7$, Mark Gurwell, $^8$, Stefanie Milam, $^9$} 

\affil{$^1$Departments of Astronomy and of Earth and Planetary Science, University of California, Berkeley, CA 94720, USA; \email{imke@berkeley.edu}}
\affil{$^2$National Radio Astronomy Observatory, Socorro, NM 87801, USA; \email{bbutler@nrao.edu}}
\affil{$^3$School of Physics, University of Melbourne, Victoria, 3010, Australia; \email{rsault@nrao.edu}}
\affil{$^4$SOFIA / USRA, NASA Ames Building N232, Moffett Field, CA 94035, USA; \email{amoullet@usra.edu}}
\affil{$^5$Department of Earth and Planetary Science, University of California, Berkeley, CA 94720, USA; \email{chris.moeckel@berkeley.edu}}
\affil{$^6$Department of Earth and Planetary Science, University of California, Berkeley, CA 94720, USA; \email{jtollefs@berkeley.edu}}
\affil{$^7$Division of Geological and Planetary Sciences, California Institute of Technology, Pasadena, CA 91125, USA; \email{dekleer@caltech.edu}}
\affil{$^8$Harvard-Smithsonian Center for Astrophysics, Cambride, MA 02478 USA; \email{mgurwell@cfa.harvard.edu}}
\affil{$^9$NASA Goddard Space Flight Center, Astrochemistry Laboratory, Greenbelt, MD 20771 USA; \email{stefanie.n.milam@nasa.gov}}

% This section is for ADS Processing.  There must be one line per author.
\paperauthor{Imke de Pater}{imke@berkeley.edu}{ORCID-0000-0002-4278-3168}{University of California}{Department of Astronomy and of Earth and Planetary Science}{Berkeley}{CA}{94720}{USA}
\paperauthor{Sample~Author2}{Author2Email@email.edu}{ORCID_Or_Blank}{Author2 Institution}{Author2 Department}{City}{State/Province}{Postal Code}{Country}
\paperauthor{Sample~Author3}{Author3Email@email.edu}{ORCID_Or_Blank}{Author3 Institution}{Author3 Department}{City}{State/Province}{Postal Code}{Country}

%This is the Astronomical Society of the Pacific (ASP) 2014 author template file.  This sample author template includes some common \LaTeXe formatting examples and the ASP author checklist.

\begin{abstract}
Radio wavelength observations of solar system bodies are a powerful method of probing many characteristics of those bodies.  From surface and subsurface, to atmospheres (including deep atmospheres of the giant planets), to rings, to the magnetosphere of Jupiter, these observations provide unique information on current state, and sometimes history, of the bodies.  The ngVLA will enable the highest sensitivity and resolution observations of this kind, with the potential to revolutionize our understanding of some of these bodies.  In this article, we present a review of state-of-the-art radio wavelength observations of a variety of bodies in our solar system, varying in size from ring particles and small near-Earth asteroids to the giant planets. Throughout the review we mention improvements for each body (or class of bodies) to be expected with the ngVLA. A simulation of a Neptune-sized object is presented in Section \ref{sim}. Section \ref{conc} provides a brief summary for each type of object, together with the type of measurements needed for all objects throughout the Solar System.    

\end{abstract}
\section{The Solar System as a Link between Protoplanetary Disks and Exoplanets} %IdP 
It is widely accepted that planets form in protoplanetary disks surrounding stars. However, the details of such processes remain vague. Several thousand exoplanets have been detected, varying in scale from larger than Jupiter-sized down to almost moon-sized, and in stellar distance from less than 0.01 AU to many AU's from the star. It is not clear, though, where these planets formed, since they may well have formed at a very different location from that where they have been detected today. In particular, it has been difficult to explain the presence of giant planets at stellar distance $<$ 1 AU, which triggered development of theories about planetary migration. Large planets may have formed further out, and subsequently migrated inwards as a consequence of angular momentum exchange with the disk. Planets could also migrate at a later stage if/when clearing the area near them of planetesimals. Planet migration is a difficult topic in planet formation theories, and hence no consensus has yet been reached on the effect of such migrations, not even for our own Solar System. Formation models for the latter may be hardest, since such models have to explain not only the present orbits of both large and small bodies, but also their composition and possible presence of satellites and rings. Composition of bodies in our Solar System provides one key element in formation models that would link protoplanetary disks to exoplanets. For example, the composition of comets and bodies in the outer Solar System is dominated by water ice, while bodies closer to the Sun have larger fractions of rock and metal. However, such bodies are not devoid of water, so how did the water get there; i.e., where did the planets and the planetesimals that make up the planets form originally?\par
As mentioned above, composition is a key parameter for planetary formation models, and forms a common thread throughout this chapter. In order to determine bulk composition of a body and/or its atmosphere, one needs to understand the entire body, i.e., the physics, dynamics and chemistry in a body's interior and atmosphere. Observations at radio wavelengths provide unique information as they probe regions inaccessible by nearly all other remote sensing techniques and wavelengths. For example, at radio wavelengths one can probe depths up to many tens of bars in atmospheres of the giant planets, and centimeters to meters below the surface of a solid body like a terrestrial planet, a tiny asteroid, or a planetary ring. Magnetic fields can be probed through emissions by electrons in such fields. In order to extract the most information, high spatial and high spectral resolution is essential, as well as  sensitivity to large scale structure. Since objects in our Solar System are highly time variable, due e.g., to plume activity (comets, satellites), volcanic eruptions, atmospheric winds, and simply  the rotation of a body, excellent sensitivity and imaging characteristics on short timescales (minutes or less) is highly desirable. These are attributes being planned for the ngVLA. To strengthen the case to build such an array in the near-future, in this chapter we highlight several cases where the ngVLA can make a difference.\par
Good prior reviews on radio observations of our Solar System are provided by \citet{butler2004} and \citet{depk2014}. We will build upon these reviews. In this chapter we will omit comets, as the ngVLA case for comets is provided in a separate chapter \citep{cordiner2018}. Radar measurements are discussed by \citet{lazio2018}.  We will also not consider observations of the Sun, as they are different enough to deserve their own treatment.  Lastly, we will not investigate the newly-approved long baselines (> 800 km) of ngVLA, as while they have some uses for solar system investigations (the aforementioned radar, and spacecraft tracking, for instance), in general there is not enough brightness temperature sensitivity to be useful for observations of solar system bodies.\par
\section{Giant Planets} %IdP
Observations of the giant planets in the frequency range of the ngVLA (1--116 GHz) are sensitive to both thermal and non-thermal emissions. These emissions are received simultaneously, and can be distinguished from each other by examination of their different spatial, polarization, time (e.g., for lightning), and spectral characteristics. Given the sensitivity and resolution of the ngVLA, detailed maps of both of these types of emissions will be possible. In the following subsections we highlight some recent results obtained with the VLA, CARMA, and ALMA, and discuss how these can be improved with the ngVLA.
\subsection{Deep Atmospheres}  %IdP
Understanding the coupling of gas abundances, temperature, and dynamics of the deep atmospheres of the giant planets is vital to our understanding of these planets as a whole, and to our understanding of extrasolar giant planets. 
\subsubsection{Radio Spectra}
The atmospheres of the giant planets all emit thermal (blackbody) radiation. At radio wavelengths sources of opacity are collision induced absorption (CIA) by hydrogen, and absorption by NH$_3$, H$_2$S, PH$_3$, and H$_2$O gases (opacity by clouds can probably be ignored; \citep{depater2018}). For near-solar composition atmospheres\footnote{For a solar-composition atmosphere we adopt the proto-solar values of \citet{asplund2009}: C/H$_2 = 5.90 \times 10^{-4}$; N/H$_2= 1.48 \times 10^{-4}$; O/H$_2= 1.07 \times 10^{-3}$; S/H$_2= 2.89 \times 10^{-5}$; Ar/H$_2= 5.51 \times 10^{-6}$.}, most of the atmospheric opacity has been attributed to ammonia gas, which has a broad absorption band near 22 GHz. Indeed, for Jupiter and Saturn NH$_3$ gas indeed dominates, but not for Uranus and Neptune. The S/N ratio on these planets appears to be considerably enhanced above the solar value, and spectra can be modeled well by including opacity from H$_2$S gas, and selectively increase the H$_2$S abundance considerably (factors of $\sim$~10 and 30 for Uranus and Neptune, resp.) (e.g., \citep{depmitch1993}, and references therein). Example spectra for Jupiter and Neptune are shown in Fig.~\ref{fig:spec}.\par
\articlefigure{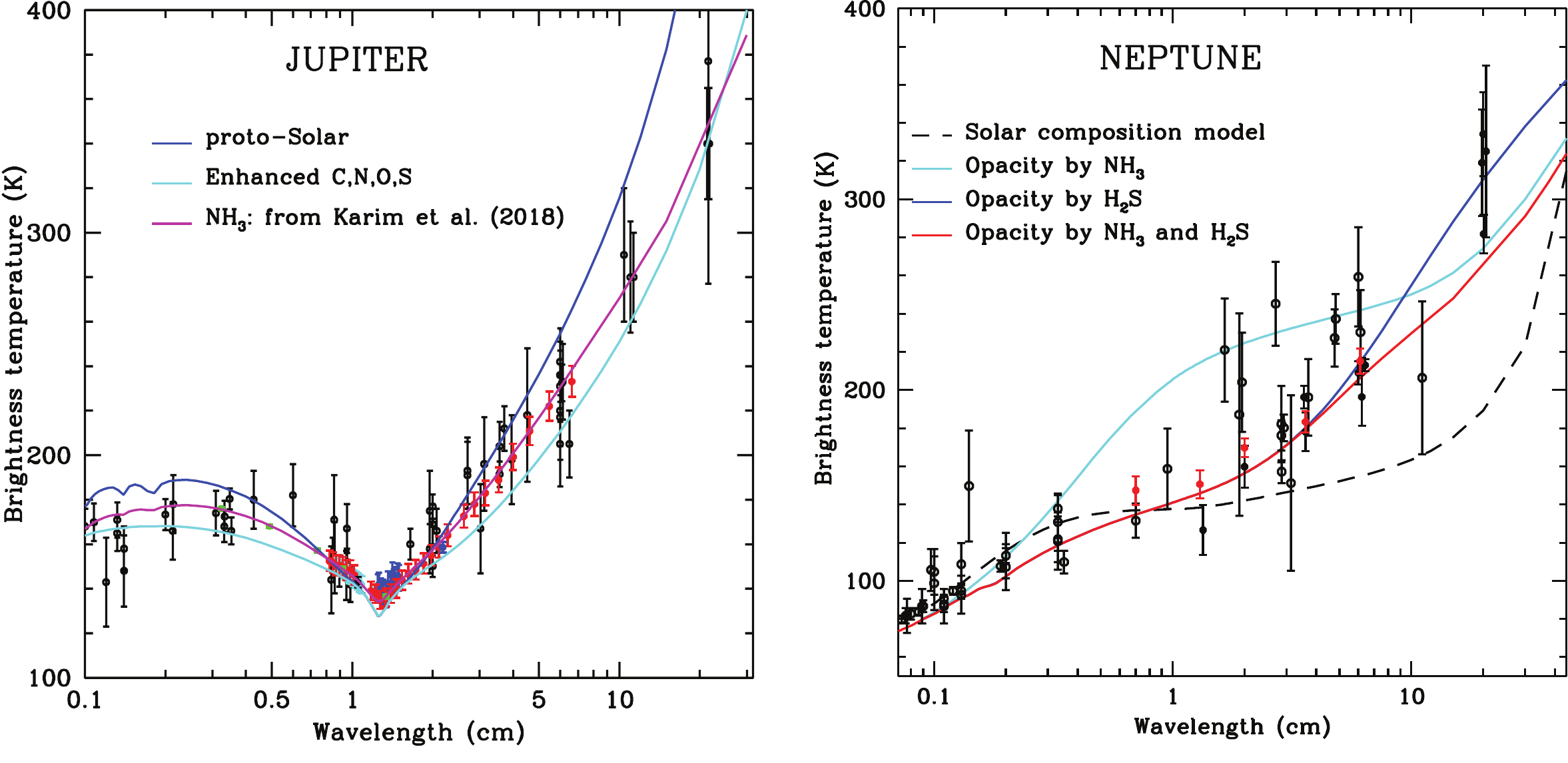}{fig:spec}{Disk-averaged spectra for Jupiter and Neptune, with various models superposed. Jupiter: The red data points are recent VLA values. Various model calculations are superposed: blue: solar abundances for all gases; cyan:  NH$_3$ and H$_2$S are enhanced by a factor of 3.2 compared to the solar values, and H$_2$O and CH$_4$ by a factor of 4; magenta: \citet{karim2018} decreased NH$_3$ decreased compared to the cyan curve to match the data at 0.5--2 cm. Figure is adapted from \citet{depater2018}; for more clarifications, the reader is referred to that paper. Neptune: The red data points are VLA values from 2003 \citep{depater2014}. The curves are model calculations, as indicated. Opacity by all gases was included in the red curve. In the cyan curve, opacity due to H$_2$S was ignored, and in the black curve absorption by NH$_3$ was ignored.
}
% Fig_spec.pdf is the PDF file; fig:spec is the label, i.e., in the main text call fig as: \ref{fig:spec}
%
\subsubsection{Radio Maps}
 The thermal emission from all four giant planets has been imaged with the VLA. To construct high signal-to-noise images, the observations need to be integrated over several hours, resulting in maps that are smeared in longitude and only reveal brightness variations in latitude. In order to discern longitudinal structures on radio maps, such as the Great Red Spot (GRS) on Jupiter, \citet{sault2004} developed an algorithm to essentially take out a planet's rotation. This algorithm has recently been applied to data of all planets obtained with the VLA after its upgrade. Below we highlight a few results.\par
{\it Jupiter.} A longitude-smeared image of Jupiter is shown in Fig.~\ref{fig:jup}a. This composite multi-wavelength radio image of Jupiter reveals numerous bright bands across the disk. These bands are roughly co-located with the brown belts seen at visible wavelengths. A longitude-resolved map is shown in panel b. The image shows an incredible amount of detailed structure, including the GRS and Oval BA, but also numerous small vortices, hot spots and plumes. The observed variations have been attributed to spatial variations in NH$_3$ gas, caused by a combination of atmospheric dynamics and condensation at higher altitudes \citep{depater2016}.\par 
Spectra of zones and belts from the longitude-smeared images, and spectra of individual features on the longitude-resolved maps have been used together with radiative transfer calculations to determine the altitude distribution of ammonia gas at these locations. These calculations show that ammonia gas is brought up from the deep atmosphere to the cloud condensation levels in the plumes (and equatorial zone), and that the dry air is descending in the hot spots and North Equatorial Belt down to 20 bar or deeper. The hot spots coincide with hot spots seen at a wavelength of 5 $\mu$m; the plumes have been hypothesized to form the counterpart of the equatorial Rossby wave \citep{depater2016} theorized to produce the 5 $\mu$m hot spots (e.g. \citep{showman2004}).\par
\articlefigure{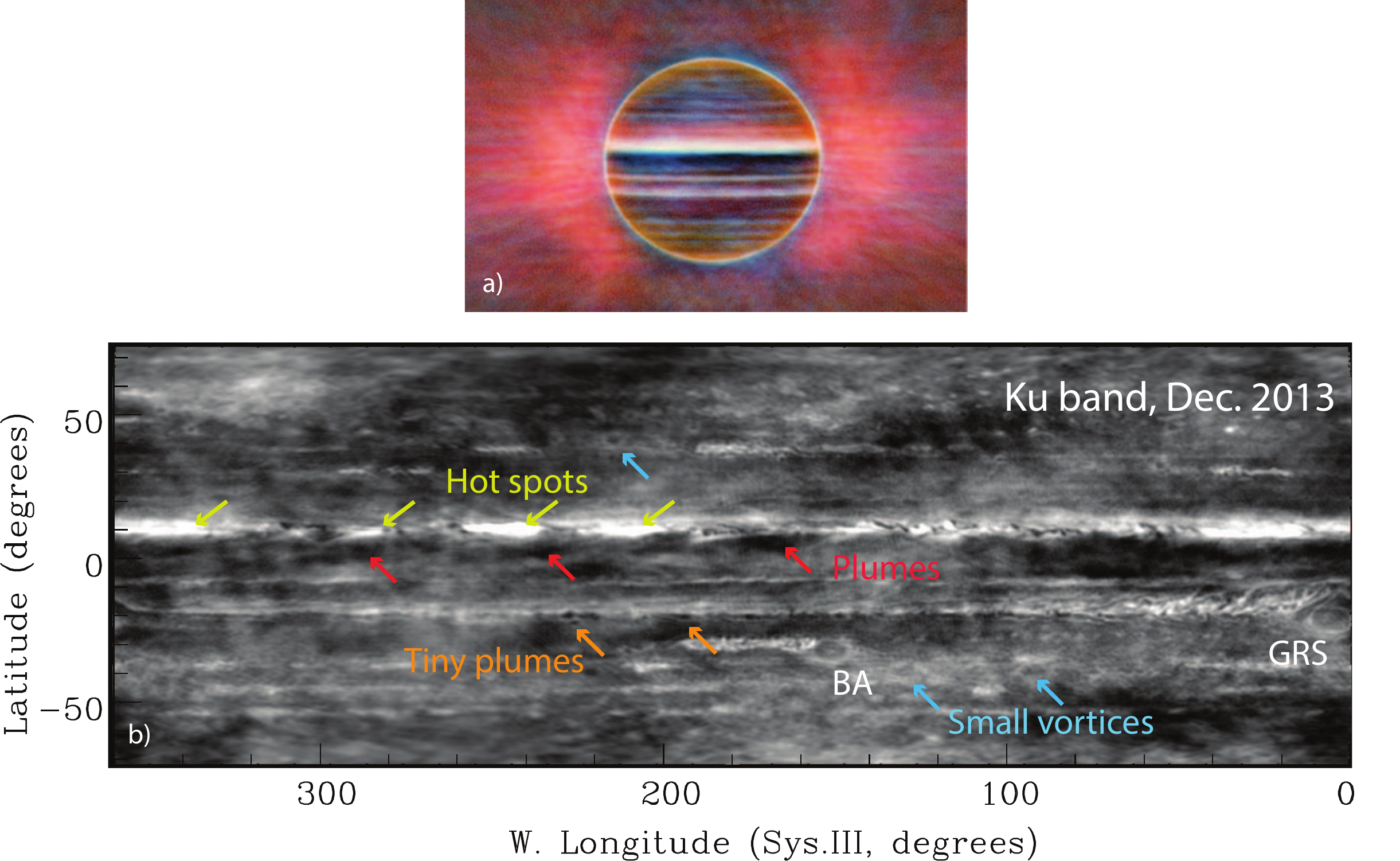}{fig:jup}{a) Radio image of Jupiter constructed from VLA data taken between December 2013 and May 2014 at three wavelengths: 2 cm in blue, 3.5 cm in gold, and 6 cm in red. A uniform disk had been subtracted to better show the fine banded structure on the planet. The pink glow surrounding the planet is synchrotron radiation produced by spiraling electrons trapped in Jupiter's magnetic field. This image is averaged from several 10-hr observing sessions, so any longitudinal structure is smeared by the planet's rotation. \citep{depater2016}. b) Longitude-resolved map of Jupiter at a wavelength of 2 cm (Ku band, 12-18 GHz), from panel a). As in panel a), bright features indicate a high brightness temperature, or low NH$_3$ abundance, so deeper warmer layers are probed. Much finescale structure can be discerned. The Great Red Spot (GRS) and Oval BA are indicated, as well as hot spots (yellow arrows), ammonia plumes (red), small vortices (cyan), and tiny ammonia plumes (orange). (Adapted from \citep{depater2018})}
{\it Saturn.} Images of Saturn's microwave emission reveal, in addition to the planet itself, its rings. Using data from the upgraded VLA, Fig.~\ref{fig:ZZ} shows the rings and planet in exquisite detail. The planet itself is visible through its thermal emission, and displays zones and belts as on Jupiter. Surrounding the north pole one can discern a hexagonal pattern, which has been seen at many other wavelengths, and both by the Voyager (1981) and Cassini (2004--2017) spacecraft. This pattern is the meandering path of an atmospheric jet stream. The hexagon itself remains nearly stationary in the rotating frame of Saturn, and has been interpreted as a westward (retrograde) propagating Rossby wave \citep{allison1990}.\par 
The emission from the planet's rings is dominated by Saturn's thermal radiation reflected off the ring particles. Only a small fraction of the radiation at centimeter wavelengths is thermal emission from the rings themselves. 
Water ice comprises the bulk of Saturn's rings, yet it is the small fraction of non-icy material that is key in revealing clues about the system's origin and age. Using the Monte Carlo Simrings package \citep{dunn2002} to fit multi-wavelength (0.7--13 cm) VLA and 2-cm Cassini/RADAR data of the rings, \citet{zhang2017,zhang2018} show that the non-icy fraction of the rings varies from 0.1--0.5\% in the B ring, to 1--2\% in the C ring, and that the particles overall are quite porous (75\%-90\%, depending on location in the rings). They further showed that there is a band in the middle C ring where the intrinsic thermal emission is almost constant with wavelength, and which has an anomalously high non-icy material fraction (6--11\%),  This has been interpreted by the presence of large particles, composed of rocky cores covered by porous, icy mantles. Assuming that the non-icy fraction is due to continuous impacts by micrometeorites, the rings have been estimated to be no older than 200 Myr, while the middle C ring might have been hit by a rocky Centaur 10--20 Myr ago.\par
\articlefigure{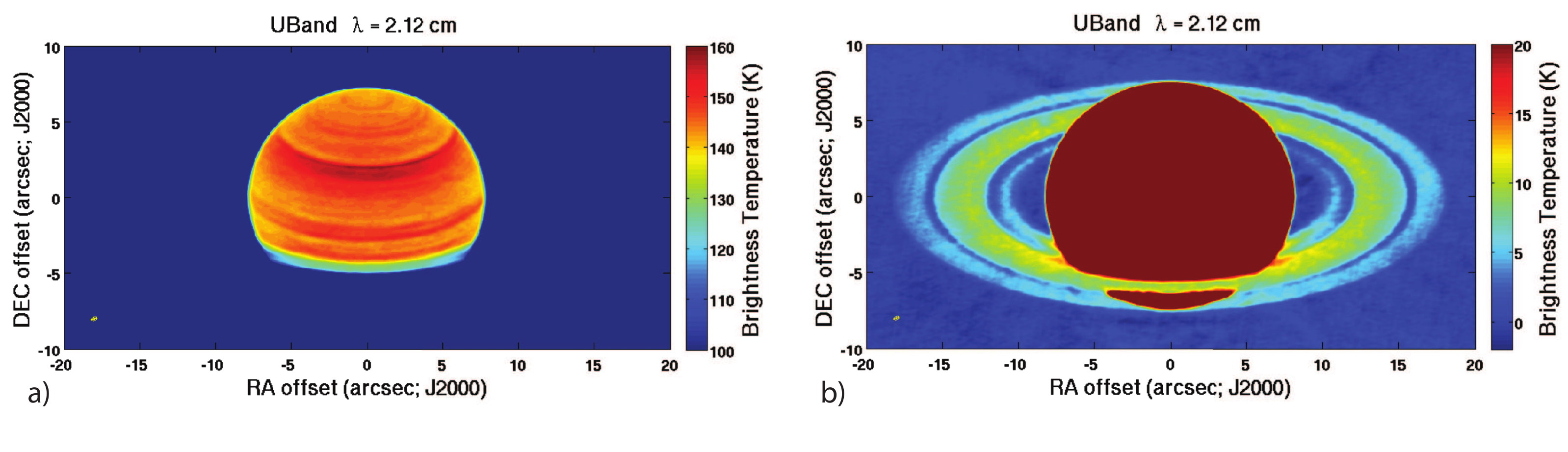}{fig:ZZ}{a) Radio image of Saturn at a wavelength of 2 cm (Ku band, 12--18 GHz). The color range is set from 100K to 160K to highlight Saturn's radio bands. b) The same image as in panel a), but here the  color is set from -3 K to 20 K to show the structure of the rings. The yellow ellipses on the lower left of each panel indicate the synthesized beam sizes and corresponding position angles. \citep{zhang2018}}
A longitude-resolved image of Saturn is shown in Fig.~\ref{fig:satmap}b. Since the observations were obtained over time slots of only 2 hours, we did not cover a full rotation of the planet at each wavelength. As shown, by combining the X and Ku band data we could cover most of a rotation.
%In Fig.~\ref{fig:satmap}b we show a map of Saturn constructed from data obtained with the Cassini spacecraft at 15 GHz. 
A comparison with similarly obtained Cassini maps between 2009 and 2011 \citep{janssen2013} shows that the bright turbulent band at 30--40$^\circ$N is caused by the giant storm system at that latitude in 2010-2011 (Fig.~\ref{fig:satmap}a), which at the time caused that band to be anomalously bright at a wavelength of 2 cm. Even in May 2015 that band is still anomalously bright at radio wavelengths.\par
\articlefigure{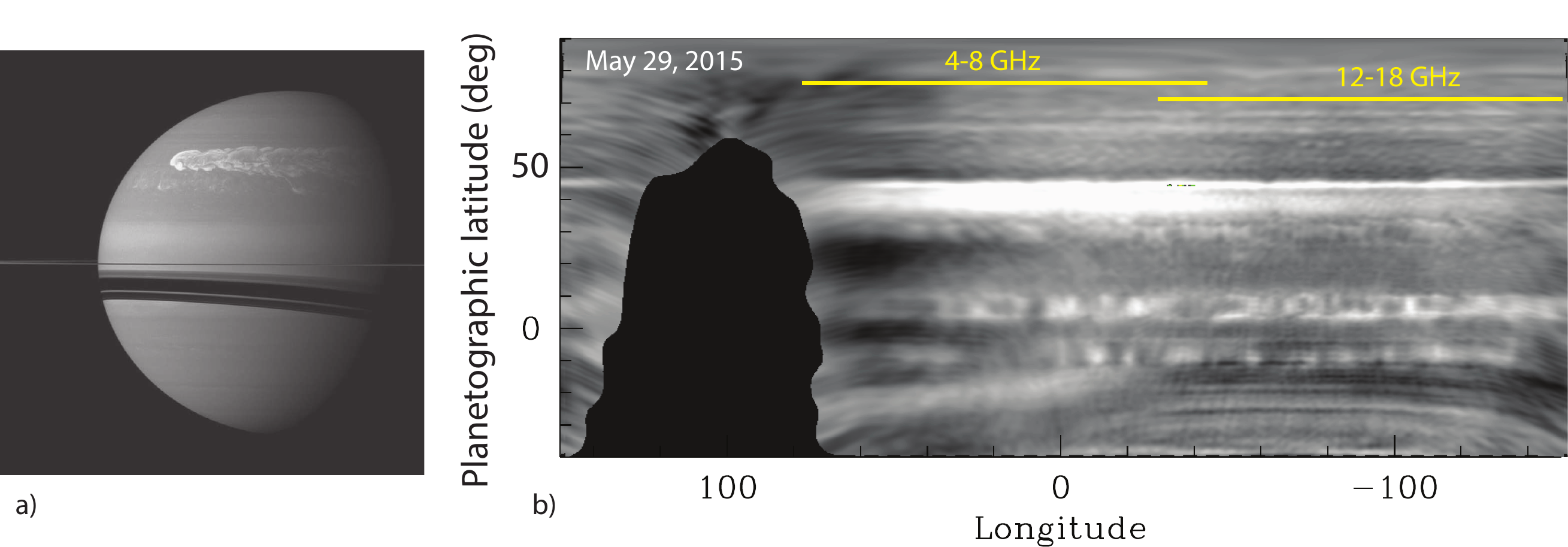}{fig:satmap}{a) Image of Saturn taken by the Cassini spacecraft on 25 February 2011 about 12 weeks after a powerful storm was first detected in Saturn's northern hemisphere. This storm is seen overtaking itself as it encircles the entire planet. (NASA/JPL/Space Science Institute, PIA12826).
b) VLA longitude-resolved map from May 2015, after subtraction of a uniform disk. As in Fig.~\ref{fig:jup}b, bight features indicate a high brightness temperature, or low NH$_3$ abundance.
Observations at X (8-12 GHz) and Ku (12-18 GHz) were combined; together they cover almost a full rotation of the planet. (Sault \& de Pater)
%Cylindrical map of Saturn's 2-cm brightness temperature constructed from Cassini radiometer observations taken in March 2011. The value plotted is the residual brightness relative to a model for a fully saturated atmosphere. The black stripe across the equator is Saturn's ring, blocking the planet's thermal radiation from behind. The planet was mapped by continuous pole-to-pole scans during 14 h when the spacecraft was near periapse. Periapse (indicated by the dashed line) was at 3.72 Saturn radii, where the resolution was best (1.6$^\circ$ in latitude). The resolution degrades linearly with spacecraft distance, out to 5.64 Saturn radii. \citep{janssen2013}.
}
{\it Uranus.} Radio observations of Uranus obtained between the early 60's and 80's revealed a steady increase in its disk-averaged brightness temperature, which was attributed to the poles potentially being warmer than its equator, a hypothesis confirmed using the imaging capabilities of the VLA (e.g., \citet{depater1988}); several images are shown in Fig.~\ref{fig:ur}, which clearly show that both poles are much brighter than the equator. Both the 2005 and 2015 images show several broad bands on Uranus. Uranus pole has an intriguing structure as well, though not dissimilar to that seen at other poles: a bright polar dot, surrounded by a slightly darker (cooler) region, encircled by a bright ring, like the polar collar seen at infrared wavelengths. Longitude-resolved images using the technique from \citet{sault2004} revealed no longitudinal structure.\par
ALMA images at 3 mm show a similar banded structure, and these data reveal Uranus's $\epsilon$ ring as well.\par 
\articlefigure{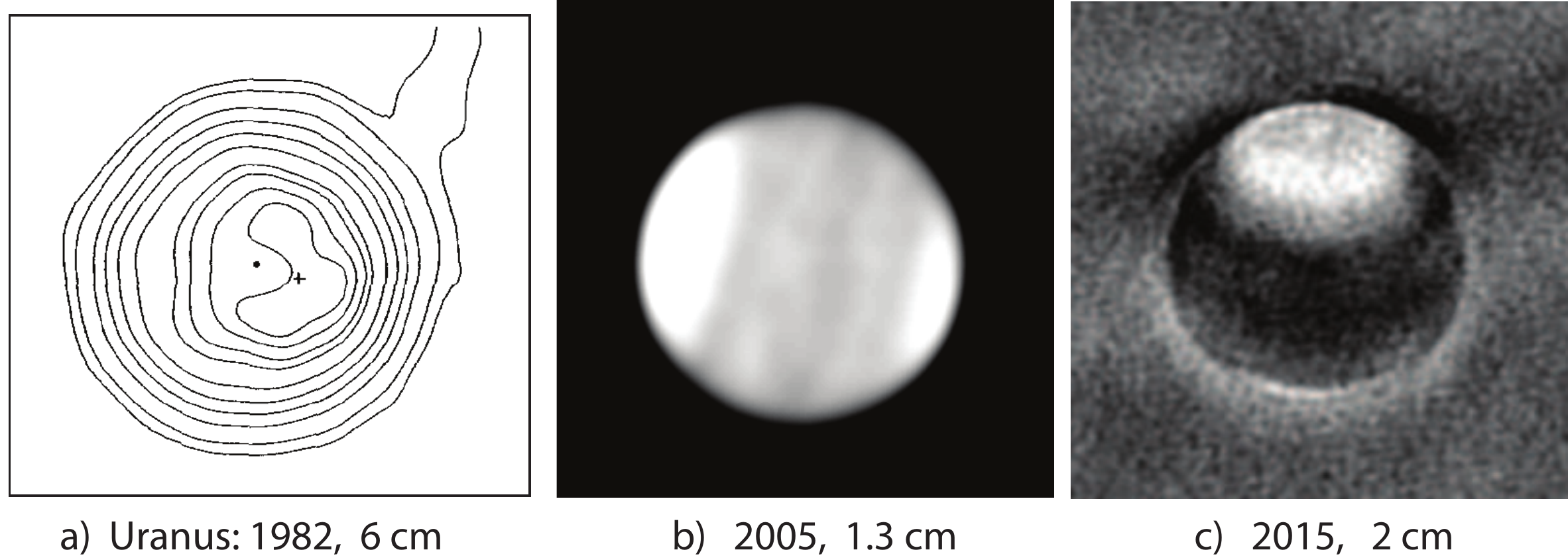}{fig:ur}{Radio images of Uranus taken at in different years and at different wavelengths: 1982 at 6 cm, when the south pole (indicated by a +) was near the subsolar point (the dot) \citep{depater1988}. b) 2005 at a wavelength of 1.3 cm, (Hofstadter \& Butler). c) 2015, a preliminary image at 2 cm (de Pater \& Sault). In all images the planet is roughly 3.5--4" across. All maps were constructed from many hours of data, so longitudinal features are smeared by the planet's rotation. The bright regions near the poles imply a relative lack of absorbing gases (H$_2$S, NH$_3$) above both poles.}
{\it Neptune.} Multi-wavelength observations obtained in 2003 showed that Neptune's south pole is considerably enhanced in brightness at both mid-infrared and radio wavelengths, i.e., from $\sim$~0.1 mbar levels in the stratosphere down to tens of bars in the troposphere \citep{depater2014}. The enhanced brightness observed at mid-infrared wavelengths is interpreted to be due to adiabatic heating by compression in the stratosphere, and the enhanced brightness temperature at radio wavelengths reveals that the subsiding air over the pole is very dry; at altitudes above the NH$_4$SH cloud at $\sim$~40 bar it is only $\sim$~5\% compared to that at deeper layers or other latitudes. This low humidity region extends from the south pole down to latitudes of 66$^\circ$S, the latitude of the planet's south polar prograde jet. Based upon the combination of near-- and mid-IR and radio data, the authors suggested a global circulation pattern where air is rising above southern and northern midlatitudes, from the troposphere up well into the stratosphere, and subsidence of dry air over the pole and equator. Although this model is corroborated by more recent VLA data (Fig.~\ref{fig:nep}a), it does not explain all observations (see, e.g., \citep{tollefson2018}).\par 
Fig.~\ref{fig:nep}b shows a longitude-resolved map in Ku band. No obvious longitudinal structure can be discerned, despite the fact that there was a large storm system in the south that was connected to a dark vortex, features expected to be rooted in the deep atmosphere. This feature should be located near 340$^\circ$ W. longitude and 30--40$^\circ$S latitude. Alhough we do see a dark, i.e., cold, area potentially caused by rising gas, this region does not stand out over other dark regions at that latitude.\par
\articlefigure{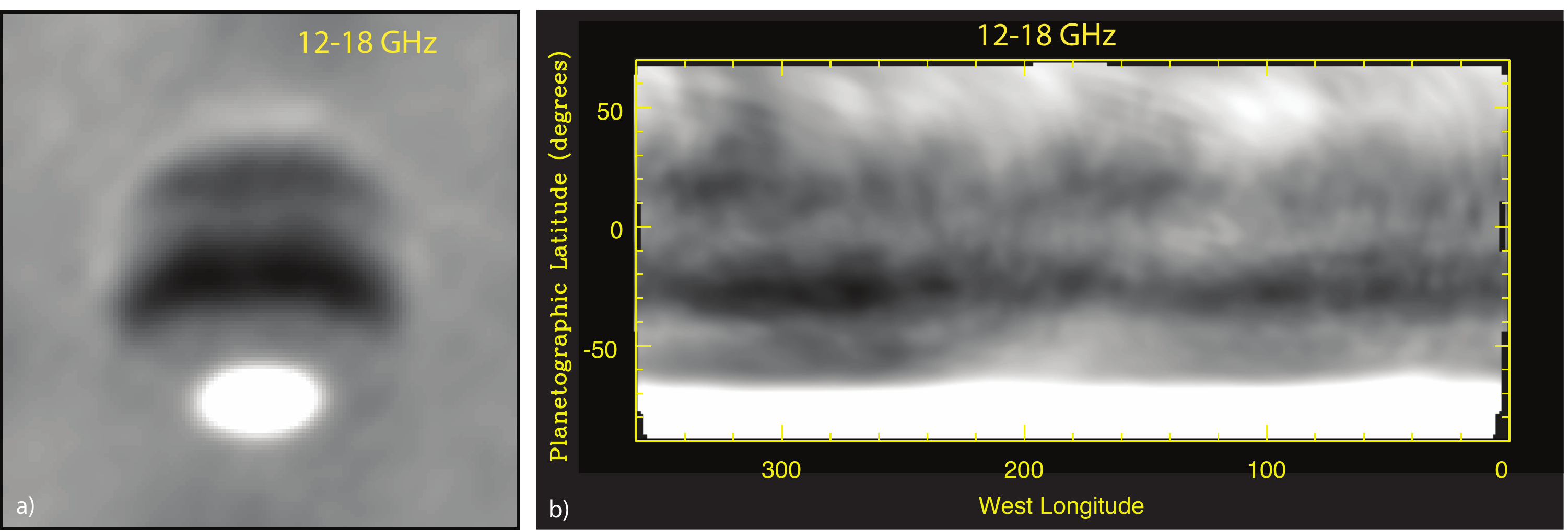}{fig:nep}{a) Longitude-smeared image of Neptune in Ku (2 cm; 12--18 GHz) band, taken in 2015. A uniform limb-darkened disk has been subtracted from the data. Brighter areas indicate higher brightness temperatures, likely caused by a low opacity (H$_2$S gas), whereas darker areas indicate cold areas. b) Longitude-resolved map from the same data. The undulating banding pattern may be caused by artefacts. (Sault \& de Pater)} 
\subsection{Stratosphere} %IdP will start
Several molecules are prevalent in the stratospheres of the giant planets, such as hydrocarbons C$_m$H$_n$ (formed e.g., via photochemistry from methane gas), H$_2$O, CO and HCN (resulting from infalling materials, and/or brought up from the deep atmosphere). The distribution and line profiles of these species will help distinguish between species (or fractions thereof) brought up from the deep atmospheres versus those brought in from the outside. The data can also be used to derive the wind profiles as a function of latitude and depth.\par  
Prominent emission lines of CO and HCN were detected on Neptune in the early 1990s \citep{marten1993}, with abundances $\sim$~1000 times higher than predicted from thermochemical models. While HCN cannot have been brought up as such from Neptune's deep interior (it would condense), it is still unclear whether a fraction of the CO in Neptune's stratosphere is brought up from below, or if all of it has an external origin. Since the production of CO in Neptune's interior depends on the water abundance (CH$_4$ $+$ H$_2$O $\rightarrow$ CO $+$ H$_2$) and the vertical mixing rate, an accurate measurement of CO constrains the water abundance in the planet's interior. The CO 1-0 line is the most sensitive to a potential internal source through accurate measurements of the wings of the line, which are seen in absorption (originate in the troposphere), in contrast to the emission line at its center (originating in the stratosphere). One of the challenges is to observe the entire line, which is several GHz wide, with a narrow emission peak $\lesssim$~10 MHz.
\subsection{Synchrotron radiation} %IdP, CM
Synchrotron radiation has only been detected from Jupiter. It is emitted by high energy ($\sim$~1--100 MeV) electrons trapped in the planet's radiation belts. The synchrotron emission morphology and intensity is changing over time due to the planet's rotation, as well as in response to impacts (e.g., the impact of comet Shoemaker-Levy 9; \citep{depater1995}), changes in the solar wind ram pressure, and any other phenomena (internal or external to the magnetosphere) that induce changes in the energetic electron distribution in Jupiter's magnetosphere.\par 
Jupiter's synchrotron radiation has been imaged at frequencies between 74 MHz and 22 GHz (e.g., \citet{depk2014}, and references therein), usually in all 4 Stokes parameters, which helps to better constrain the models (e.g., magnetic field field geometry). A recent VLA image of the planet's radio emission at 20 cm is shown in Fig.~\ref{fig:jupL}. Because the radio emission is optically thin, and Jupiter rotates in 10 h, one can use tomographic techniques to map the 3D radio emissivity, assuming the emissions are stable over 10 h. The shape of Jupiter's radio spectrum is determined by the intrinsic spectrum of the synchrotron radiating electrons, the spatial distribution of the electrons and Jupiter's magnetic field. There are substantial variations over time in the spectrum. Changes in the radio spectrum most likely reflect a change in either the spatial or intrinsic energy distribution of the electrons. With the ngVLA we may begin investigating the cause of such variability through its imaging capabilities at high angular resolution at different wavelengths (quasi)-simultaneously, while at the same time being sensitive to short spacings.\par 
Synchrotron radiation has not been detected from any of the other giant planets; a sensitive search might reveal such radiation, or put stringent constraints on such emissions.\par
\articlefigure{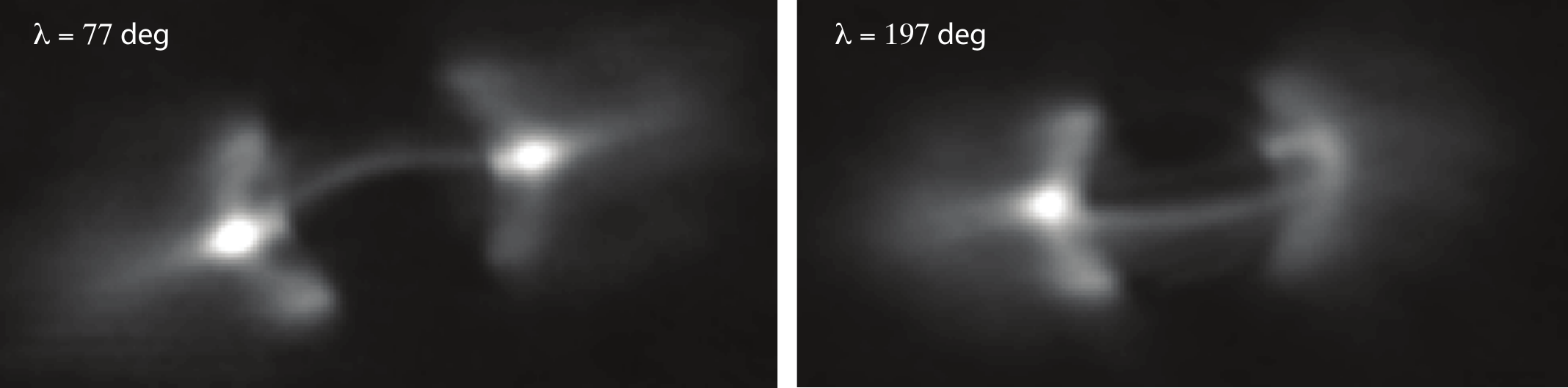}{fig:jupL}{Jupiter's synchrotron radiation at two different viewing aspects (indicated by the longitude $\lambda$ in Jupiter's System III coordinate system) as observed with the VLA in Jan. 2014. The thermal emission (limbdarkend disk) was subtracted from the data so only synchrotron radiation is visible. (de Pater \& Sault)}
\subsection{Outlook for the ngVLA}
With an instrument that is 10 $\times$ more sensitive and has a 30 $\times$ higher spatial resolution, maps of the giant planets should be superb. We show a simulation and comparison with the VLA and ALMA in Section~\ref{sim}. 
\section{Terrestrial Planets} %(BB) (IdP started it, based upon copy/paste previous papers)
Radio wavelength observations of the terrestrial planets (Mercury, Venus, the Moon and Mars) are important tools for determining atmospheric, surface and subsurface properties. For atmospheric studies, such observations can help determine temperature, composition and dynamics, similar to observations of the giant planets discussed in Sections 2.1 and 2.2. For surface and subsurface studies, such observations can help determine temperature, layering, thermal and electrical properties, and texture.\par 
Thermal emission from a body's surface can be described as greybody Planck radiation, departing from the blackbody radiation spectrum by a wavelength-dependent factor representing the ability of a surface to re-emit radiation at a given wavelength (spectral emissivity). Since surfaces are not opaque at those wavelengths, 'surface' thermal emission originates in part below the surface, and needs to be determined by performing full radiative transfer modeling including vertical variations of soil properties. The strongest contribution comes from the electrical skin depth which corresponds to the level at which opacity reaches $\sim$1. A commonly used rule of thumb is that observations mostly probe 10-20 wavelengths deep into the subsurface, for typical soil properties.\par
Although thermal emission is intrinsically unpolarized, it becomes partially linearly polarized as it passes through the surface because of the disconinuity in the dielectric constant at the surface-space (or atmosphere) interface. As observed from Earth, the polarization is zero at the center, and increases towards a maximum value at the limb, which depends on the dielectric constant and the surface roughness. 
The 3-D temperature distribution of a body is itself determined by its orbital and rotational properties and topography (which determines local insolation), and surface thermal and radiative properties such as thermal conductivity, thermal inertia, dielectric constant, bolometric Bond albedo - which are indicative of composition and physical nature (such as porosity and roughness).\par
{\it Mercury}. A prime example of the thermal (surface) emission displaying the importance of the history of solar insolation and topography is shown by the 3.6-cm radio image of Mercury displayed in Fig.~\ref{fig:merc}a, probing a typical depth of $\sim$~70 cm. Two "hot" regions are visible, one of them almost opposite to the direction of the Sun. This hot-cold pattern results from Mercury's 3/2 spin-orbit resonance combined with its large orbital eccentricity \citep{mitchell1994}. While the surface temperature responds almost instantaneously to changes in illumination, the subsurface layers do not, and this variation in solar insolation remains imprinted at depths well below the surface, as shown in the figure. \par
Models of the emission revealed that the surface was largely basalt-free, as later confirmed by observations with the {\it MESSENGER} spacecraft. The map in Fig.~\ref{fig:merc}b shows the difference between the map in panel a) and the thermal model. The negative (blue) temperatures near the poles and along the terminator are indicative of areas colder than predicted in the model, likely caused by surface topography, which causes permanent shadowing in craters at high latitudes and transient effects in the equatorial regions, where crater floors and hillsides are alternately in shadow and sunlight as the day progresses.\par 
\articlefigure{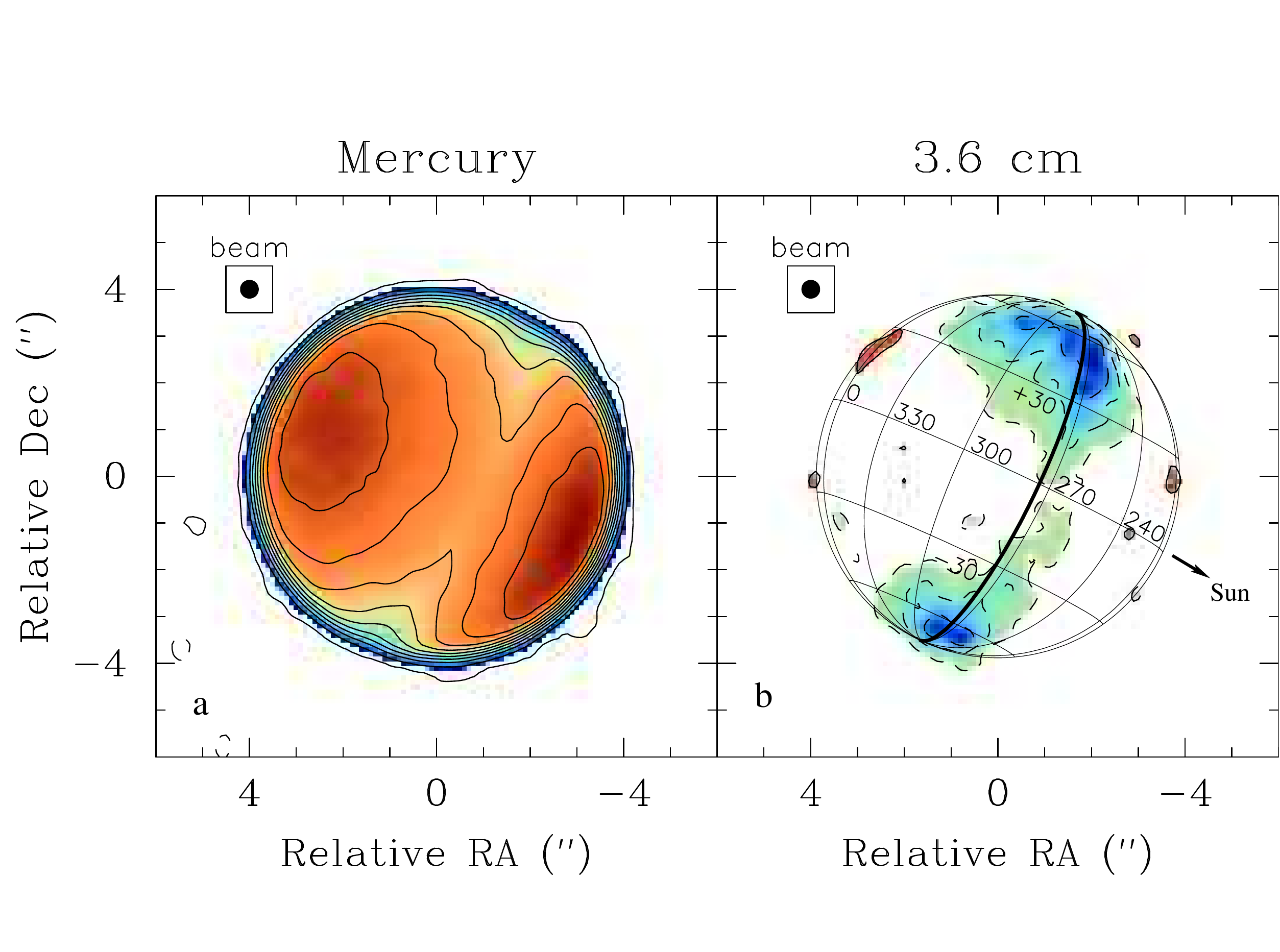}{fig:merc}{a) The 3.6-cm thermal emission from Mercury observed with the VLA. The beam size is 0.4" or 1/10 of a mercurian radius. Note the two "hot" regions, discussed in the text. b) A residual map after subtracting a model image from the map in panel a. Mercury's geometry is indicated, including the direction to the Sun and the morning terminator (heavy line). The hot regions have been modeled well, since they do not show up in this residual map. Instead, large negative (blue) temperatures near the poles and along the morning side of the terminator are visible. \citep{mitchell1994} }
{\it Venus}. Venus presents a different thermal forcing scenario than any of the other terrestrial planets.  With a thick atmosphere ($\sim$100 bars at the surface), solar insolation does not penetrate directly to the surface, except in very restricted wavelength ranges.  The thick atmosphere acts like a thermal blanket, keeping the surface at nearly constant temperature as a function of latitude and longitude.  This is moderated by surface topography, of course, where temperatures are much cooler at the tops of the mountains.  Previous observations, including those from the {\it Magellan} spacecraft, have allowed for a detailed understanding of the composition of the atmosphere and top layers of the surface/subsurface - a model can be made which is a good match to data from mm to short cm wavelengths \citep{butler2001}.  However, it is now established that the model breaks down at longer wavelengths \citep{butler2003,mohan2017}.  A good physical explanation for that breakdown is still lacking, however, and high resolution, high sensitivity images, as could be made by ngVLA, could help address this lack of understanding.\par
The atmosphere of Venus is also a rich area of study.  With complex dynamics and chemistry, there is a long history of studying its properties at radio wavelengths (see  \citet{encrenaz2015,moullet2012} for recent examples).  Since ngVLA will have access to the CO(1-0) transition, doppler wind observations will yield much information about the distribution of that molecule as a function of latitude and longitude (local time of day).  In addition, continuum absorption by sulfur-bearing molecules allows lower frequency observations to be used to determine constraints on their abundances \citep{butler2001,jenkins2002}.  It is known from Venus Express that the abundance of these molecules is variable both spatially and temporally \citep{vandaele2017}, and to understand this over long timescales (when {\it Venus Express} will have completed its mission), we need Earth-based observations like those that can be provided by the ngVLA.\par
{\it Mars}. By contrast, the atmosphere of Mars is four orders of magnitude less abundant than that of Venus, but no less interesting.  In fact, understanding the distribution of water on the surface and subsurface of that planet (liquid or frozen) necessarily requires an understanding of the water in the atmosphere as well, and understanding water on Mars is one of the important scientific questions in the solar system.  While there have been extensive spacecraft observations of the water (and other molecules) in the atmosphere of Mars \citep{
mcconnochie2018,wolkenberg2018,montmessin2017,clancy2017}, there is no guarantee that spacecraft will continue to be sent to the planet to make such measurements.  Ground-based observations of atmospheric water are needed to provide a continuous record of the abundance of water vapor in the martian atmosphere.  The ground-state transition of water vapor near 22 GHz is observable in this way, though sufficient resolution and sensitivity are required.  The VLA has been used for such observations \citep{clancy1992,butler2005}.  While the line cannot be measured across the entire disk, there is enough path length at the limb to yield a detectable signal.  Figure~\ref{fig:marswater} shows observations of the Mars atmospheric water vapor with the VLA in 2003.\par
\articlefigure{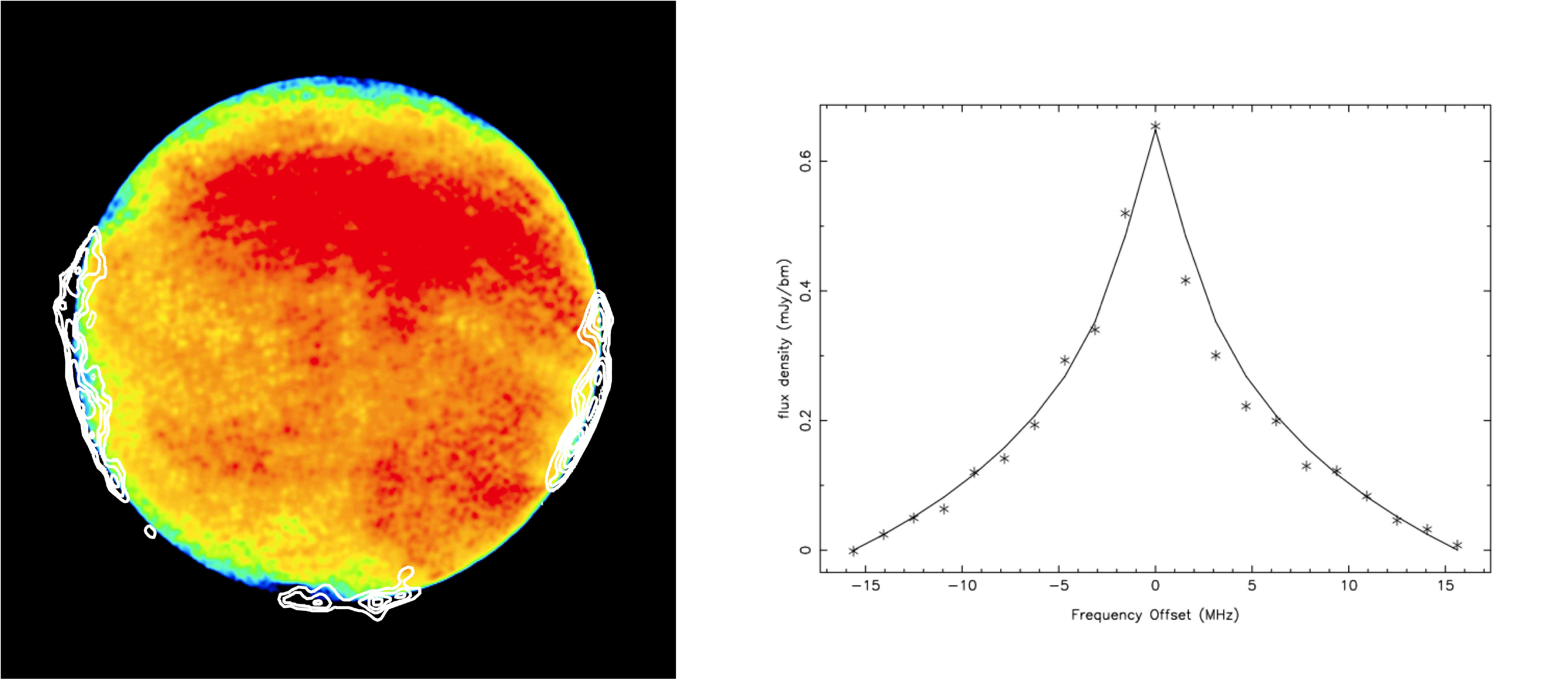}{fig:marswater}{(a) Water vapor distribution around the limb of Mars observed in 2003 with the VLA.  The colored background is continuum (thermal) emission from the surface and subsurface of the planet, with hottest regions red, through yellow, green, blue, to black for the coldest.  Note the colder south polar regions to the lower left (the image has not been rotated so that North is exactly up).  Integrated water vapor line emission is shown in white contours.  Note the lack of water vapor at northern latitudes, and the concentration at the south pole.  (b) Water vapor line shape, averaged around the limb (stars) along with a best-fit model (solid line).  Total precipitable water vapor ranges from 0 to 20 $\mu$m. \citep{butler2005}
}
\section{Satellites} 
The outer solar system's major satellites all host rocky and/or icy surfaces. While Titan is unique in our Solar System as the only moon with a dense ($\sim$~1.5 bar) atmosphere, the other major satellites have, at best, extremely tenuous atmospheres. Continuum thermal emission arising from the satellites' 35-150 K surfaces and subsurfaces dominates their signal at radio frequencies, while emission lines atop this continuum provide a fingerprint of molecular species in their atmospheres. 
\subsection{Titan} %(MG)
Titan's atmosphere is dominated by N$_2$, with CH$_4$ (at a few \% level) the only other major species. %\citep{niemann2010}, 
Like Earth, it has a well-defined troposphere and stratosphere (e.g. %\citet{lindal1983} and 
\citet{fulchignoni2005}). By mass, $\sim$90\% of the atmosphere lies below the tropopause. 
While {\it Cassini}/CIRS and its INMS system have made extensive observations of Titan's stratosphere and above, the troposphere and surface are difficult to sense due to hazes and molecular absorption (N$_2$ CIA and CH$_4$). These regions can only be probed at a few specific infrared wavelengths (away from strong CH$_4$ absorption bands) and at radio wavelengths. Yet the troposphere and surface are regions of high interest, and since the arrival of {\it Cassini} with its cloud penetrating Ku-band RADAR, one of consistent wonder. Surface conditions (pressure, temperature) are near the triple point of methane, like water on Earth, allowing liquid methane on the surface and gaseous methane in the atmosphere. {\it Cassini}/RADAR data revealed numerous lakes and seas \citep{stofan2007} confirming persistent liquid on its surface. The surface is shaped by aeolian and fluvial processes, and methane rain is strongly suspected from rapid equatorial changes \citep{turtle2011}.\par 
The tropopause is an ineffective cold trap for CH$_4$; it diffuses upward \citep{niemann2010} where it contributes to photochemistry and haze production. In fact, the troposphere contains the bulk of Titan's methane, not the surface lakes, and the stability of the methane cycle requires replacement over geologic timescales \citep{horst2017}. The methane cycle is thus tightly tied to the temperature of the lower atmosphere, which has only been measured via {\it Voyager 1} \citep{lindal1983} and {\it Cassini} radio occultations \citep{schinder2012}, and during the {\it Huygens probe} descent \citep{fulchignoni2005}.
Thus the lower atmosphere is poorly measured, despite holding most of the atmospheric mass. Broad latitudinal temperature seasonal variability, at the level of several K, is seen at the tropopause \citep{anderson2014} and the surface \citep{jennings2016, legall2016}, but the situation within the troposphere is not well sampled, while this information is crucial for climate models \citep{tokano2014}.\par 
{\it Cassini}'s 13.8 GHz (Ku-band) RADAR and radiometry data show that Titan's surface displays large scale variations in brightness, due in part to variations in surface roughness (enhancing scattering) and in absorptivity (dielectric constant) of the surface materials. The reflectivity of Xanadu, an extended and peculiar low latitude region, is anomalously high, indicative of coherent scattering on scales of a few wavelengths. In contrast, the dunes and plains which cover most of Titan are radar-dark, with an effective (low) dielectric constant that is indicative of a cover of organic material precipitated from the atmosphere. How these differences vary with seasonal or longer timescales is unknown. While {\it Cassini} leaves a rich scientific legacy, it only covered half of a Titan year; continued investigation of longterm climate and seasonal variability must rely on new facilities and methods.\par
Broadband observations of Titan at cm, mm, and submm wavelengths are affected primarily by CIA from N$_2$-N$_2$ and N$_2$-CH$_4$, with the strength scaling roughly as $\nu^2$. Thermal emission from Titan at frequencies below 40-50 GHz mostly originates from the (sub)surface, while higher frequencies are mostly sensitive to the atmosphere; with increasing frequency, peak sensitivity rises in altitude. Thus sensitive broadband observations spanning the cm-mm bands allow for sounding the troposphere from the (sub)surface to near the tropopause, providing a full characterization of the troposphere and surface brightness temperature. The ngVLA, combining sensitivity and linear resolution gains of order 10 over the current VLA, will allow high precision surface brightness temperature maps for exploration of surface limb darkening, localized emissivity, and latitudinal temperature gradients, with sensitivity in a few hours to under 0.3K in a 35 mas beam at Ka band (225 km on Titan, the size of moderate geologic regions). At Ku band the sensitivity is superb, allowing comparison of the brightness temperatures of the leading and trailing hemispheres, and a direct high resolution tie to the {\it Cassini} 13.8 GHz RADAR and radiometry data. Adding higher frequencies, the ngVLA will allow detection of latitudinal and/or regional variations in tropospheric temperature, or potentially methane abundance. In the 3mm band the stratosphere can be probed via observations of several nitriles (e.g., HCN, HC$_5$N, HC$_7$N), providing information on the stratospheric chemistry and dynamics. 
\subsection{(Sub)-surface Structure of Other Satellites} % (KdK)
As for the terrestrial planets and Titan, in the 1-100 GHz range observations of satellites are sensitive to regions from the upper few cm down to meters within the subsurface and provide quantitative information on temperature and material properties at these depths.\par 
%Measurements at multiple frequencies capture the vertical thermal profile over this subsurface region.  The surface temperature and vertical thermal profile are set by albedo, time of day, and the thermal properties of the material, and thermal measurements therefore constrain these properties as a function of depth. \par
%
The galilean satellites' disk-integrated brightness temperatures have been measured with single-dish telescopes and arrays including the VLA from 5-230 GHz \citep{ulich1984,muhleman1991,depater1984,butler2012}. The observations of Ganymede and Europa require a drop in emissivity from near 1 at 100 GHz to closer to 0.5 at 15 GHz, with the latter emissivity consistent with that of cold ice. However, Callisto's high brightness temperature across this frequency range deviates from this pattern and remains a mystery. More recently, spatially-resolved observations with the IRAM interferometer and ALMA have provided brightness temperature maps of the galilean satellite subsurfaces at millimeter wavelengths, some of which are shown in Figure~\ref{fig:galsatsurf} \citep{moullet2008,dekleer2018}. The maps demonstrate the expected correlations with surface albedo, but also reveal localized regions with temperatures and thermal properties distinct from the surrounding area.\par
While ALMA has provided unprecedented spatial resolution at $\sim$cm depths in the galilean satellite subsurfaces, lower frequencies are needed to access the deeper (tens of cm to meters) subsurface regions. Multi-frequency studies utilizing both ALMA and the ngVLA could provide 3D thermal maps of the subsurfaces of the galilean satellites, yielding the first detailed information on the subsurface material properties over a large depth range, and associate specific thermal properties with surface compositional units. The improved sensitivity of the ngVLA will also enable brightness temperature measurements at these frequencies of satellites farther out in the Solar System, while the improved spatial resolution will aid in detecting satellites in close angular proximity to their primaries. Observations of multiple satellite systems over a range of frequencies would enable comparisons between the surfaces of icy satellites at different solar distances.\par
\articlefigure{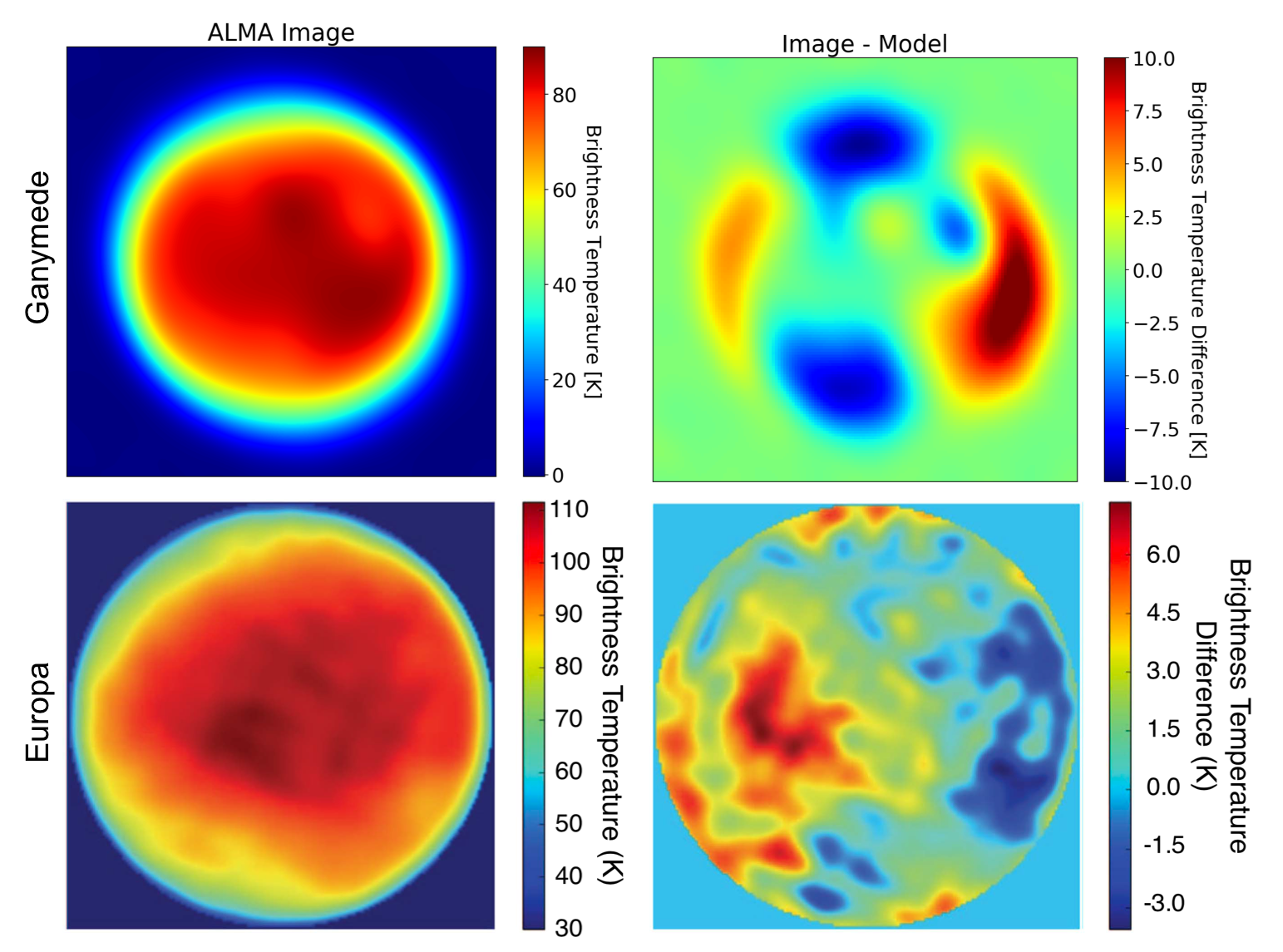}{fig:galsatsurf}{ALMA continuum images of Ganymede and Europa at 230 GHz before and after subtraction of a model. The Ganymede model is a simple disk model and the difference image demonstrates surface temperature variations on the order of 10 K, which largely correlate with surface albedo \citep{dekleer2018}. The Europa observations were made at a higher spatial resolution, and the residual image is based on a thermal model, demonstrating that such models can account for much but not all of the surface temperature variations \citep{trumbo2017}. While ALMA frequencies are sensitive to the $\sim$cm subsurface, ngVLA observations of these satellites would probe deeper regions.}
\subsection{Atmospheres \& Plumes of Satellites other than Titan} % (KdK)
The tenuous atmospheres of the outer Solar System's satellites (excluding Titan) are sourced by the sublimation and sputtering of surface ices, with a contribution from geological activity in the case of volcanic or cryovolcanic moons. These processes generate atmospheres that vary spatially, dictated by the distribution of surface frost, pattern of particle bombardment, or locations of gas plumes. Orbital and seasonal changes in particle environment and solar illumination drive variability on a range of timescales. Investigations into the nature and variability of thin satellite atmospheres inform our understanding of the creation and evolution of surface conditions on these worlds, and shed light on their potential habitability. \par
Radio frequencies trace the composition and dynamics of thin satellite atmospheres through rotational transitions of molecular species. Data that resolve these narrow lines have the potential to constrain the atmospheric temperature and its diurnal and seasonal variations, which is a largely unexplored arena. High-sensitivity data may detect new species, or improve our understanding of the gas sources through the simultaneous measurement of multiple gas species. Long-baseline arrays with high sensitivity can even map the distribution of gas species and constrain their sources through spatial correlations with properties such as the surface frost distribution, bombardment pattern, or geological formations. Doppler shift measurements allow for wind speed mapping and characterization of atmospheric dynamics. Io's sulfur-based atmosphere is the densest ($\sim$nbar) and best-studied of the jovian satellites. A variety of gas species have been detected and mapped in Io's bulk atmosphere at millimeter wavelengths, including SO$_2$, $^{34}$SO$_2$, and SO \citep{moullet2010,moullet2013}, and the SO$_2$ emissions have been used to map wind speeds \citep{nowling2015}. The ngVLA will provide a tool that extends the high sensitivity and spatial resolution of ALMA down to lower frequencies, enabling searches for new species as well as mapping of a wider range of lines for known species, improving gas temperature estimates. \par
On satellites with volcanic or cryovolcanic activity, radio observations are directly sensitive to molecular gases within the plumes themselves. Radio-frequency studies of these active worlds have the potential to detect plumes and determine their composition, providing direct information on the geophysical processes driving the activity as well as the properties of the source reservoir (water or magma). Using ALMA, \citet{moullet2015} have mapped alkali gases in Io's volcanic plumes (see Figure \ref{fig:galsatatm}); the data resolve the distribution of these gases into distinct localized areas. With the recent end of the \textit{Cassini} mission, in-system observations of the Enceladus plume have now ended, hindering efforts to determine the dependence of plume strength on Enceladus' long-period (multi-year) orbital evolution \citep{ingersoll2017}. Enceladus' plume is dominated by water, but molecules such as ammonia, methanol, formaldehyde, and even more complex organics have been detected \citep{postberg2018}. Radio frequencies are uniquely sensitive to these molecules, and spectral line observations  present a potential avenue for continuing to track the plume's variability, as well as for detecting and determining the abundance of species in the plumes of Enceladus and of other satellites that may have ongoing plume activity such as Triton and Europa.\par %
\articlefigure{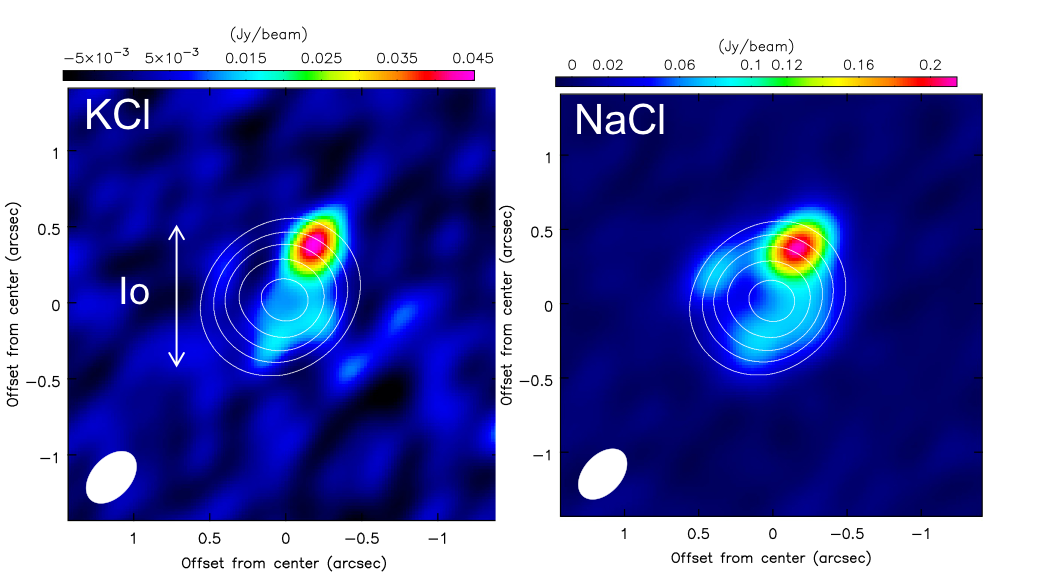}{fig:galsatatm}{Spatial inhomogeneities in gas species are easily seen in observations that resolve the disk, as shown in this example of spatially-resolved ALMA maps of alkalis in Io's volcanic plumes. The data were taken in June 2015, at a spatial resolution of 0.25"$\times$0.4". (Moullet) }
\section{Minor Bodies}% (AM)
Observations of small Solar System bodies are an essential component of planetary studies at large. On one hand, characterization of individual bodies (including composition of surfaces, interiors and -in some cases - atmospheres) provides insight into processes affecting surfaces, as demonstrated by the extensive observations of dwarf planets Ceres and Pluto by the {\it Dawn} and {\it New Horizons} spacecraft, respectively. On the other hand, global population studies on the distribution of orbital, compositional and physical properties can bring clues pertaining to the thermal, dynamical and collisional history of the early Solar System (see e.g., \citet{demeo2014}), for example constraining planetary migration and the origin of some of the giant planets' moons. These investigations are especially relevant for objects orbiting in the outer Solar System, such as Trans-Neptunian Objects (TNOs), which are considered to be some of the most pristine remnants of the protoplanetary disk, and akin to debris disks observed in other planetary systems.\par
As mentioned in Section 3, the thermal emission from a body's (sub)surface depends on a large number of variables. Several thermal models have been developed to interpret thermal emission from asteroids and TNOs, and allow to classify bodies' thermal behavior based on the value of a single parameter, the beaming factor $\eta$, representing the effective thermal inertia (e.g., \citet{lagerros1996}). In general, a low thermal inertia is indicative of a powdery nature while high thermal inertia can be linked to rock-like surfaces. Large-scale studies have shown that asteroids tend to display a much smaller effective thermal inertia than TNOs \citep{delbo2015}, which suggests that regolith depth (and overall surface processing) decreases with distance to the Sun. However relatively few measurements are available for TNOs (observations are challenging due to large distances and low temperatures), and more sources must be observed to lower observational biases and firmly establish correlations between properties and differences across populations.\par
Thermal emission mapping, especially combined with monitoring on relevant timescales (diurnal/ seasonal), provides the most informative thermal data as it allows one to disentangle the effects of local hour and latitude from geographically-tied features. Since only a few dozens of objects are larger than 0.1" in apparent size, this technique has rarely been performed.
In the absence of maps, disk-averaged thermal observations combined with optical photometry allows one to retrieve the bolometric Bond albedo and effective diameter through the radiometric method \citep{lebofsky1989}. While the results from this method depend on the assumed thermal model, it is the most effective technique to estimate sizes of a large number of otherwise spatially unresolved bodies, which can be used to deduce the collisional history for different populations of small bodies \citep{schlichting2011}. In addition, combination of thermal photometry at different wavelengths is essential to adjust thermal models for a given body, further improving the robustness of models \citep{brownbutler2018}. \par
ngVLA's sensitivity is necessary to obtain cm-wave measurements with a SNR better to what can be obtained with th VLA or ALMA, giving access to objects with smaller apparent sizes. The specificity of cm-wavelengths lies in the ability to access deeper depths - down to a meter below the surface and beyond the diurnal skin depth. The combination of ngVLA measurements across the full spectral range with other thermal measurements is essential to capture the complete thermal structure in the upper meter of surface material, and determine variations of properties with depth which can be related to ongoing processing, differentiation or formation history. An excellent spatial resolution would surpass the ability of ALMA to detect thermal spatial features (including at depths not probed by ALMA), on smaller targets, and to separate the thermal emission from contact binary systems, which are relatively frequent amongst TNOs.\par
In addition to measuring continuum thermal emission, ngVLA's spectral coverage is also adequate to target the OH 18-cm maser emission, which has been regularly observed in comets \citep{crovisier2002}. OH is of particular interest as it has a relatively long lifetime and is the photodissociation product of water, and hence can be used to derive a comet's water outgassing rate. It should be noted that this OH line corresponds to a solar-pumped maser transition, for which excitation strongly depends on its heliocentric velocity. Water atmospheres are not expected in the outer Solar System, but water outgassing has been suggested for Ceres, and for some objects of the Centaur class. OH detection would provide a strong proof of the reality of transient outgassing events, which would carry significant implications for the composition and surface processes on those bodies.\par
\section{Simulations}\label{sim} %(Moeckel, Tollefson)
To highlight the capabilities of the ngVLA, we have constructed a toy model of a planet with typical features: banded structure indicating zonal features, and gaussian features such as upwelling plumes and downwelling hotspots. We used Neptune as a sample case and superimposed structure on the planet. The features are constructed to be smallest at the limb with increasing size and spacing towards the center of the planet. The simulation was done for various ngVLA configurations: using antennas just in the core, using the core and the spiral arms extending outwards to 10 km (plains) and using the full 1000km array (revB). Furthermore, we simulate our toy Neptune models at 100 GHz and compare the resulting sensitivities and resolutions to the extended VLA-A and ALMA configurations. All simulations were created in CASA with simobserve using an integration time of 60s and a total time of 1200s (similar results are obtained with an integration time of only a few minutes). This creates a visibility file from which we create an image with CASA tclean using multiscale and natural weighting. The results for the different configurations are shown in Figure~\ref{fig:sims-ngvla}. Our toy model (panel a) is shown after subtraction of a limb-darkened disk (disk-averaged brightness temperature $\sim$~120 K); it is populated with both bright and dark bands and spots, that have a contrast of $\sim 1\%$ of the background, or 1-2K. Panel b) shows the sky as seen by the compact inner array (plains). The lower resolution leads to a smearing of many of these features, however, the large hotspots can still be seen protruding through the bands due to the very high sensitivity. Panel c) shows the  structure imaged with the full array, where features that are 1K in magnitude and about 0.01" in size can be distinguished.\par
Panels d-f) shows the toymodel from panel a) as it would be seen using the extended configurations for ALMA (d), the VLA (e), and the ngVLA (f; which is equal to c). This comparison highlights the scale of features we could detect, where table~\ref{table:simulations} outlines the resulting beam size and RMS for each configuration. The banded structure can be seen across all simulations, where the VLA substantially smears the bands and cannot pick up the fine-scale structure at the limb. ALMA can pick up the banded structure relatively well, however, fails at resolving the point sources in the upper right quadrant of the planet. Only the ngVLA is able to pick up a substantial number of the small point sources, including features as small as 0.01 arcsecs. This can be generalized to resolving structures down to a size of order $<10$km/AU, and with that an order of magnitude better than the VLA and Juno in the case of Jupiter. In terms of sensitivity, the large collecting area of the ngVLA results in the lowest RMS in the beam, as seen by the lack of structure off the disk.\par
\articlefigure{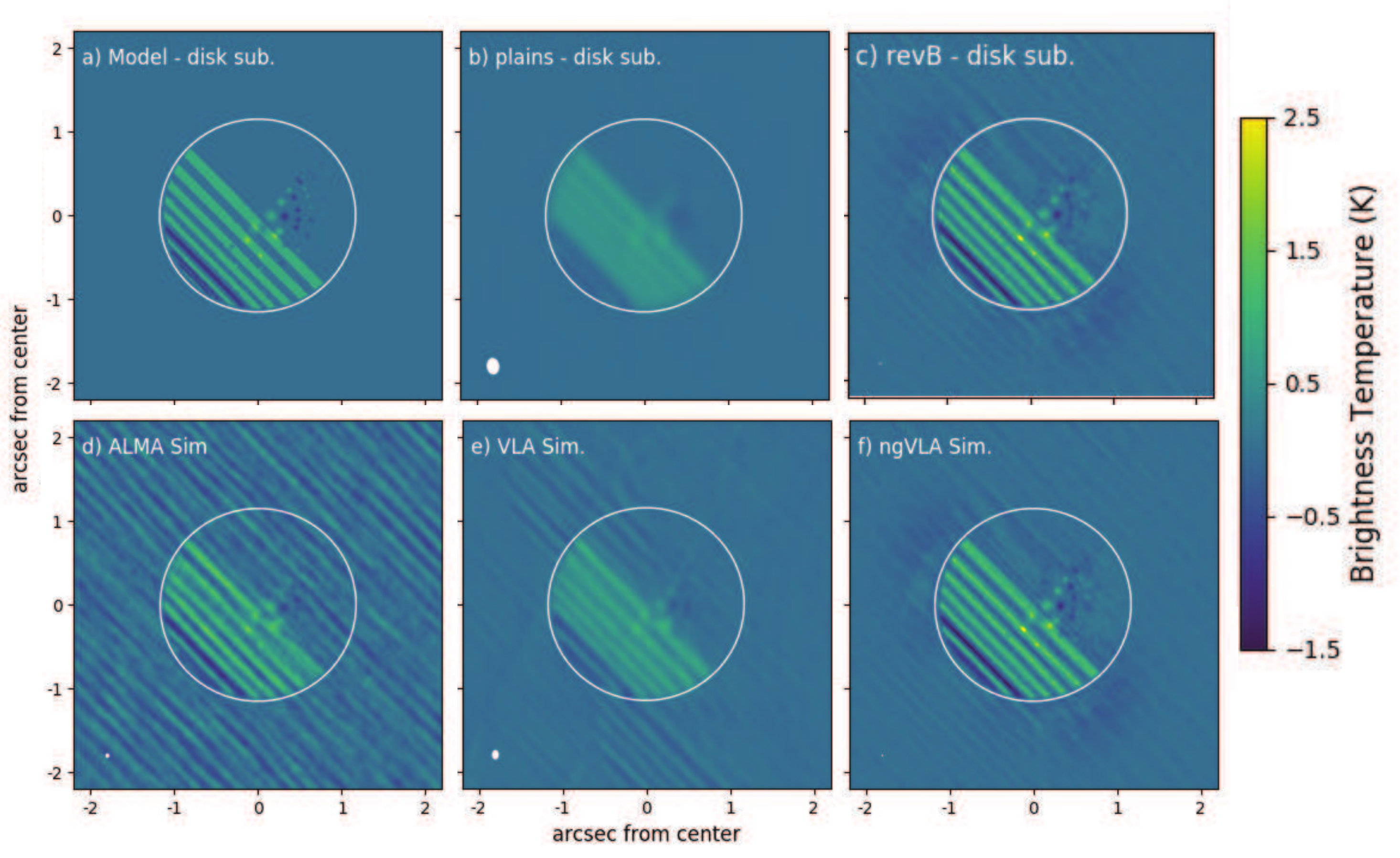}{fig:sims-ngvla}{Model ngVLA simulations for various configurations of a toy Neptune outlined by the white disk. The brightness temperature of the various features is of order 1-2 K. The beam size is shown as a white oval in the bottom left of each simulation. a) The original sky model, after subtraction of a limb-darkened disk. Note the positive (bright) and negative (dark) stripes in the southern hemisphere and spots near the center, and extending up into the northern hemisphere up to $\sim$~1/4 of a radius from the northern limb. b) Model as seen by the compact inner array (plains). c) Model as seen using the fully extended array (revB). 
d) Toymodel as seen by ALMA in its most extended configuration. e) Toymodel as seen with the VLA in its most extended (A) configuration. f) Same as in panel c. (Moeckel \& Tollefson)}
\begin{table}
\caption{Information about the model simulations}
\begin{tabular}{l l l l l}
\\
\hline
Image & Configuration file & Beam Size & RMS (K) & tclean niter \\ \hline
ALMA sim & alma.cycle5.10.cfg & 0.05''x0.05'' & 0.20 & 1000 \\
VLA sim & vla.a.cfg & 0.11''x0.08'' & 0.03 & 5000 \\
ngVLA sim & ngvla-revB.cfg & 0.02''x0.02'' & 0.02 & 15000 \\
\hline
\label{table:simulations}
\end{tabular}
\end{table}
\section{Conclusion and Recommendations}\label{conc}% (IdP,BB)

$\qquad$ $\bullet$ {\it Giant Planet Atmospheres:} To determine the composition and unravel the dynamics in planetary atmospheres we need to map the 3D distribution of gases in the troposphere (e.g., NH$_3$, H$_2$S) through mapping at continuum wavelengths across the entire ngVLA band at high spatial resolution ($\sim$~0.01"), and by mapping the stratosphere of the planets at particular molecular line transitions (e.g., CO, HCN, H$_2$O) at high spatial and spectral resolution ($\sim$~0.01" and $\sim$~100 kHz) to spectrally resolve the emission core of the lines and spatially the limb of the planets.   

$\bullet$ {\it Giant Planet Synchrotron Radiation:} We distinguish two goals: 1) Map the 3D distribution of Jupiter's synchrotron radiation from 0.3-30 GHz. 2) Search for synchrotron radiation from the other giant planets. Measurements of all four Stokes parameters are key to interpret the data in terms of magnetic field geometry and electron (spatial $+$ energy) distributions. 

$\bullet$ {\it Planetary Rings}: High precision maps of rings at high spatial resolution and at different frequencies (2-100 GHz) are needed to determine the thermal and scattered (planet) light emission from the rings, which provides information on e.g., the particle size distribution and composition of rings. The mass fraction of non-icy material in Saturn's rings provides an estimate of the age of the ring system. Polarization will help interpret the data.

$\bullet$ {\it Terrestrial Planets}: Atmospheres and the (sub)surface structure can be studied in detail (composition, structure, dynamics) with high spatial resolution maps across the entire ngVLA band, in total intensity and, for surfaces, in linear polarization. Both low and high spectral resolution is required to study molecular line transitions throughout a planet's troposphere and stratosphere.

$\bullet$ {\it Satellites}: To determine the composition and dynamics in their atmospheres and (sub)surface properties (e.g., dielectric constant) we need to map the satellites across the entire ngVLA band both in total intensity and linear polarization, and map the atmospheres and plumes at high spectral and spatial resolution (similar to giant planet stratospheres).  

$\bullet$ {\it Small Bodies} To understand the dynamical and physical evolution of the small body populations, ngVLA observations across the entire ngVLA band can be carried out for a large number of small bodies (asteroids, Centaurs, Kuiper-Belt objects), which provide us with the necessary fundamental surface properties such as thermal inertia, conductivity and albedo, as well as equivalent size. 

\subsection{Measurements Required}

$\qquad$ {\bf Frequency coverage}: A wide coverage in frequency  ($\lesssim$~1--116 GHz at minimum) is needed for atmospheric and (sub)surface mapping. At low frequencies, we typically probe deeper layers in atmospheres and surfaces, and shallower levels at higher (continuum) frequencies.

{\bf Field of View and Mosaicking}: For Jupiter and Saturn a field-of-view of order 2-3 arcmin is needed to map the planet, rings and Jupiter's synchrotron radiation. 
  
{\bf Angular Resolution}: Any improvement in spatial resolution with respect to the VLA would directly translate into the ability to distinguish smaller features. For most solar system studies, a high angular resolution is required (0.01"), but simultaneous coverage of large scale structures.   An angular resolution of order $\sim$~0.1" would match JWST at $\sim$~5 microns; 0.01" would match near-IR observations with future ELTs equipped with adaptive optics. 
 
{\bf Imaging}: Observations on short and long baselines are essential to capture the large range of spatial scales at play, and must be obtained simultaneously due to the relatively short timescales of atmospheric dynamics (compare to weather on Earth), volcanic eruptions (Io), plumes, and cometary outgassing. 

{\bf Sensitivity}: Since all planets rotate (Jupiter \& Saturn: $\sim$~10 hrs; Uranus \& Neptune: 16-17 hrs), and winds with typical velocities of $\sim$~100 m/s will move features with respect to each other, one ideally would like to get instantaneous high SNR. With the VLA we observe over a full rotation of the planet, and "derotate" the planet in the UV plane so as to create a longitude-resolved map of a planet (assuming features do not move with respect to each other over 10-20 hrs. For ngVLA we should ideally be able to image an object within a min of time. For Jupiter, a 5-min integration leads to a smearing of 1". Typical desired sensitivities are of order 0.1 K.

{\bf Frequency Switching}: Due to the fast timescales involved in studies of time-variable phenomena, it is essential to obtain observations at different frequencies essentially simultaneously, i.e., by being able to switch frequencies often and fast.

{\bf Flexible Scheduling Capabilities, Cross Coordination, \& ToO}: Solar System observations often require simultaneous observations at different telescopes. For example, studying a giant planet's atmosphere, simultaneous data at mid-IR wavelengths can help determine the temperature structure in an atmosphere, a key parameter to model radio spectra. Data at near-IR, optical and UV wavelengths provide information on aerosols and hazes, while the 5-$\mu$m wavelength range probes deeper layers in the atmospheres of Jupiter and Saturn, providing information complementary to that at radio wavelengths. It is therefore important to schedule observations simultaneously with other telescopes (e.g., HST, JWST, 8-10 m telescopes, and ELTs in the future). ToO observations are extremely important as well, such as a quick response to impacts on a planet, a comet or near-Earth asteroid apparition, or a volcanic eruption.

{\bf Tracking of Objects}: All objects in the solar system require non-sideral tracking. In order to track objects such as near-Earth asteroids and satellites orbiting a (itself moving) planet, a minimum rate of 1"/s is required.

{\bf Polarization}: Measurements of all 4 Stokes parameters are highly desired, as these contribute much to our understanding of Jupiter's synchrotron radiation, scattering off planetary rings, and thermal (sub)surface emissions.

\subsection{Synergies with HST, JWST, ALMA, SKA, LOFAR}% (IdP, SM)

Complementary information can be obtained by simultaneously observing at the ngVLA with, e.g., HST/JWST, ALMA, SKA, and/or LOFAR.
For example, the deep atmospheres (below the optical cloud deck) of giant planets can only be observed at radio wavelengths, although for some planets (Jupiter, Saturn) one can probe down to ~7 bar in cloud-free regions at 5 $\mu$m wavelength. Complementary information is obtained at 5 $\mu$m and radio wavelengths (e.g., \citet{sault2004}), which would be useful to better understand the atmospheres' composition and dynamics. Ideally high spectral resolution measurements in the 5-$\mu$m range are obtained, so that the abundance profiles of molecules such at PH$_3$, NH$_3$, H$_2$O, and CH$_3$D can be determined \citep{bjoraker2015}. JWST will cover visible, near- and mid-IR wavelengths at superb sensitivity, though not necessarily at a high spatial nor spectral resolution. Visible and near-IR wavelength observations provide information on clouds and storm regions in planetary atmospheres; with ngVLA one can relate these features to the deeper atmospheric layers, and probe, perhaps, the cause of such systems (caveat: the observations need to be conducted simultaneously!). At mid-IR wavelengths the temperature structure of the atmosphere in the upper troposphere and stratosphere can be determined. Combining observations at all different wavelengths has the potential to unravel the 3D temperature and composition throughout the atmosphere. \par
SKA will be limited to lower frequencies and shorter baselines; even if the highest frequency is 15 GHz for that telescope, the resolution will not be near to what ngVLA can achieve. LOFAR covers still lower frequencies, down to $\sim$~10 MHz. These low frequencies are interesting for , e.g., Jupiter's non-thermal radiation. \par
 With ALMA observations at higher frequencies can be obtained; simultaneous ngVLA and ALMA coverage of several spectral lines from planets and satellites would be useful to fully characterize the atmospheres (and at continuum wavelengths to probe just below the surface of solid bodies).

\subsection{Synergies with spacecraft} %Clipper, Juice, Dragonfly, Lucy} (SM)

A number of other synergies can be envisioned with direct planetary missions including past/current/future spacecraft, landers, and sample return missions. For the giant planets, continuous long-term studies to follow-up the extensive knowledge obtained from the {\it Cassini} mission towards Saturn and Titan should be acquired at all wavelengths.  Should the New Frontiers 4 mission concept {\it Dragonfly} to Titan be selected, complimentary studies will be necessary for global measurements for the interpretation of data acquired by the spectrometers aboard the spacecraft. Other missions such as {\it Juno}, {\it Juice}, and {\it Europa Clipper} in the Jupiter system will also benefit from long term studies of the system's magnetosphere from ngVLA. Accessibility to thermal measurements and lines such as OH maser emission will provide essential complementary data for missions to small bodies such as the New Frontiers 4 mission concept for a comet sample return from Comet 67P, the {\it Lucy} mission to asteroids, and long term studies to follow-up {\it Dawn}'s results.  

%\bibliography{editor}  % For BibTex

\begin{thebibliography}{}
\bibitem[Allison et al.(1990)]{allison1990}{Allison, M., Godfrey, D.~A., Beebe, R.~F.\ 1990, Science, 247, 1061.}
%\bibitem[ALMA Partnership et al.(2015)]{ALMA2015} {ALMA Partnership, Hunter, T.~R., Kneissl, R., et al. 2015, \apjl, 808, L2 }
\bibitem[Anderson et al.(2014)]{anderson2014}{Anderson, C.M., et al. 2014. Icarus, 243, 129}
\bibitem[Asplund et al.(2009)]{asplund2009}{Asplund, M., Grevesse, N., Sauval, A.J., Scott, P., 2009. Ann. Rev. Astron. Astrophys. 47, 481-522.}
\bibitem[Bjoraker et al.(2015)]{bjoraker2015}{Bjoraker, G.L., Wong, M.H., de Pater, I., \'Ad\'amkovics, M., 2015. \apj, 810, \# 122, 10 pp.}
\bibitem[Butler et al.(2001)]{butler2001}{Butler, B.J., Steffes, P.G., Suleiman, S.H., Kolodner, M.A., Jenkins, J.M., 2001.  Icarus 154, 226}
\bibitem[Butler \& Sault(2003)]{butler2003}{Butler, B.J., Sault, R.J., 2003.  IAUSS IE, 17B}
\bibitem[Butler et al.(2004)]{butler2004}{Butler, B.J., Campbell, D.B., de Pater, I., Gary, D.E., 2004.  {\it New Astronomy Reviews}, 48, 1511-1535.}
\bibitem[Butler et al.(2005)]{butler2005}{Butler, B.J., Johnston, J.G., Clancy, R.T., Gurwell, M.A., 2005.  BAAS, 37, 27.08}
\bibitem[Butler(2012)]{butler2012}{Butler, B., 2012. ALMA Memo 594.}
\bibitem[Brown \& Butler(2018)]{brownbutler2018}{Brown, M.~E., \& Butler, B.~J. 2018, arXiv:1801.07221 }
\bibitem[Clancy et al.(2017)]{clancy2017}{Clancy, R.T., et al. 2017, Icarus, 293, 132. }
\bibitem[Clancy et al.(1992)]{clancy1992}{Clancy, R.T., Grossman, A.W., Muhleman, D.O., 1992.  Icarus 100, 48}
\bibitem[Cordiner \& Qi(2018)]{cordiner2018}{Cordiner, M. A., Qi, C., 2018, This issue}
\bibitem[Courtin(1988)]{courtin1988}{Courtin, R., 1988, Icarus 75, 245}
\bibitem[de Kleer et al.(2018)]{dekleer2018}{de Kleer, K., Butler, B., de Pater, I., Gurwell, M., Moreno, R., Moullet, A., 2018. 49th LPSC Meeting, No. 2083, ID 2567.}
\bibitem[Crovisier et al.(2002)]{crovisier2002}{Crovisier, J., Colom, P., G{\'e}rard, E., Bockel{\'e}e-Morvan, D., Bourgois, G. 2002, \aap, 393, 1053}
\bibitem[de Pater et al.(1984)]{depater1984}{de Pater, I., Brown, R.A., Dickel, J.R., 1984. Icarus 57, 93-101.}
\bibitem[de Pater et al.(1995)]{depater1995}{de Pater, I., et al., 1995. Science, 268, 1879-1883.}
\bibitem[de Pater \& Gulkis(1988)]{depater1988}{de Pater, I., Gulkis, S., 1988. Icarus, 75, 306-323.}
\bibitem[de Pater \& Kurth(2014)]{depk2014}{de Pater, I., Kurth, W.S., 2014. The Solar System at radio wavelengths. {\it Encyclopedia of the solar system, 3ed.}, eds. T. Spohn, D. Breuer, and T.V. Johnson. pp. 1107-1132.}
\bibitem[de Pater \& Mitchell(1993)]{depmitch1993}{ de Pater, I., Mitchell, D. L., 1993. {\it J. Geophys. Res. 
Planets}, 98, 5471-5490}
\bibitem[de Pater et al.(2014)]{depater2014}{de Pater, I., et al., 2014. Icarus, 237, 211-238.}
\bibitem[de Pater et al.(2016)]{depater2016}{de Pater, I., Sault, R. J., Butler, B., DeBoer, D., Wong, M. H., 2016. Science, 352, 1198-1201.}
\bibitem[de Pater et al.(2018)]{depater2018}{de Pater, I., Sault, R. J., Wong, M. H., Fletcher, L. N., DeBoer, D., Butler, B., 2018. Icarus, submitted.}
\bibitem[Delbo et al.(2015)]{delbo2015}{Delbo, M., Mueller, M., Emery, J.~P., Rozitis, B., Capria, M.~T. 2015, Asteroids IV, 107}
\bibitem[DeMeo \& Carry(2014)]{demeo2014}{DeMeo, F.~E., Carry, B.\ 2014, \nat, 505, 629}
\bibitem[Dunn et al.(2002)]{dunn2002}{Dunn, D. E., Molnar, L. A., Fix, J. D., 2002. Icarus, 160, 132-160.}
\bibitem[Encrenaz et al.(2015)]{encrenaz2015}{Encrenaz, T., Moreno, R., Moullet, A., Lellouch, E., Fouchet, T., 2015. P\&SS, 113, 275.}
\bibitem[Fulchignoni et al.(2005)]{fulchignoni2005}{Fulchignoni, M., et al. 2005, Nature, 438, 785.}
\bibitem[H{\"o}rst(2017)]{horst2017}{H{\"o}rst, S.~M.\ 2017, J. Geophys. Res. (Planets), 122, 432.}
\bibitem[Ingersoll \& Ewald(2017)]{ingersoll2017}{Ingersoll, A.P., Ewald, S.P., 2017. Icarus 282, 260-275.}
\bibitem[Janssen et al.(2013)]{janssen2013}{Janssen, M. A., et al., 2013. Icarus, 226, 522-535.}
\bibitem[Jenkins et al.(2002)]{jenkins2002}{Jenkins, J.M., Kolodner, M.A., Butler, B.J., Suleiman, S.H., Steffes, P.G., 2002, Icarus, 158, 312.}
\bibitem[Jennings et al.(2016)]{jennings2016}{Jennings, D.~E., et al. 2016, \apj, 816, L17.}
\bibitem[Karim et al.(2018)]{karim2018}{Karim, R. L., deBoer, D., de Pater, I., Keating, G. K., 2018. Astron. J., 155, \#129, 8 pp}
\bibitem[Lagerros(1996)]{lagerros1996}{Lagerros, J.~S.~V. 1996, \aap, 310, 1011}
\bibitem[Lazio et al.(2018)]{lazio2018}{Lazio, J., et al., 2018, This issue}
\bibitem[Le Gall et al.(2016)]{legall2016}{Le Gall, A., et al. 2016,  J. Geophys. Res. (Planets), 121, 233.}
\bibitem[Lebofsky(1989)]{lebofsky1989}{Lebofsky, L.~A. 1989, Icarus, 78}
\bibitem[Lellouch et al.(2014)]{lellouch2014}{Lellouch, E., B{\'e}zard, B., Flasar, F.~M., et al. 2014, Icarus, 231, 323.}
\bibitem[Lindal et al.(1983)]{lindal1983}{Lindal, G.~F., et al. 1983, Icarus, 53, 348.}
\bibitem[Marten et al.(1993)]{marten1993}{Marten, A. et al., 1993, Astrophys. J. 406, 285.}
\bibitem[Mitchell \& de Pater(1994)]{mitchell1994}{Mitchell, D. L., de Pater, I. 1994. Icarus, 110, 2-32.}
\bibitem[Mohan et al.(2017)]{mohan2017}{Mohan, N., et al., 2017. Icarus 297, 119}
\bibitem[Moullet et al.(2008)]{moullet2008}{Moullet, A., Lellouch, E., Moreno, R., Gurwell, M.A., Moore, C., 2008. A\&A 482, 279-292.}
\bibitem[Moullet et al.(2010)]{moullet2010}{Moullet, A., Gurwell, M.A., Lellouch, E., Moreno, R., 2010. Icarus 208, 353-365.}
\bibitem[Moullet et al.(2012)]{moullet2012}{Moullet, A., Lellouch, E., Moreno, R., Gurwell, M., Sagawa, H., 2012. A\&A 546, A102.}
\bibitem[Moullet et al.(2013)]{moullet2013}{Moullet, A., Lellouch, E., Moreno, R., Gurwell, M., Black, J.H., Butler, B., 2013. ApJ 776:32 (9pp).}
\bibitem[Moullet et al.(2015)]{moullet2015}{Moullet, A., Lellouch, E., Gurwell, M., Moreno, R., Black, J., Butler, B., 2015. AAS DPS Meeting \#47, Abstract 311.31}
\bibitem[McConnochie et al.(2018)]{mcconnochie2018}{McConnochie, T.H., et al., 2018. Icarus, 307, 294.}
\bibitem[Montmessin et al.(2017)]{montmessin2017}{Montmessin, F., et al., 2017. Icarus, 297, 195.}
\bibitem[Muhleman \& Berge(1991)]{muhleman1991}{Muhleman, D.O., Berge, G.L., 1991. Icarus 92, 263-272.}
\bibitem[Niemann et al.(2010)]{niemann2010}{Niemann, H.~B., et al. 2010,  J. Geophys. Res. (Planets), 115, E12006.}
\bibitem[Nowling \& Moullet(2015)]{nowling2015}{Nowling, M., Moullet, A., 2015. AAS Meeting \#225, Abstract 137.11.} 
\bibitem[Postberg et al.(2018)]{postberg2018}{Postber, F., et al., 2018. Nature 558, 564-568.}
\bibitem[Radebaugh et al.(2010)]{radebaugh2010}{Radebaugh, J., et al. 2010, Geomorphology, 121, 122.}
\bibitem[Sault et al.(2004)]{sault2004}{Sault, R.J., Engel, C., de Pater, I., 2004. Icarus, 168, 336-343.}
\bibitem[Schinder et al.(2012)]{schinder2012}{Schinder, P.~J., et al. 2012, Icarus, 221, 1020.}
\bibitem[Schlichting \& Sari(2011)]{schlichting2011}{Schlichting, H.~E., Sari, R. 2011, \apj, 728, 68}
\bibitem[Showman \& Dowling(2004)]{showman2004}{Showman, A.P.,  Dowling, T.E., 2000. Science 289, 1737-1740.} 
\bibitem[Stofan et al.(2007)]{stofan2007}{Stofan, E.~R., et al. 2007, Nature, 445, 61.}
\bibitem[Tokano(2014)]{tokano2014}{Tokano, T.\ 2014, Icarus 231, 1}
\bibitem[Tollefson et al.(2018)]{tollefson2018}{Tollefson, J., et al., 2018. Icarus, 311, 317-339.}
\bibitem[Trumbo et al.(2017)]{trumbo2017}{Trumbo, S.K., Brown, S.E., Butler, B.J. AJ 154:148.}
\bibitem[Turtle et al.(2011)]{turtle2011}{Turtle, E.~P., et al. 2011, Science, 331, 1414.}
\bibitem[Ulich et al.(1984)]{ulich1984}{Ulich, B.L., Dickel, J.R., de Pater, I., 1984. Icarus 60, 590-598.}
\bibitem[Vandaele et al.(2017)]{vandaele2017}{Vandaele, A.C., et al. 2017, Icarus 295, 16.}
\bibitem[Vinatier et al.(2015)]{vinatier2015}{Vinatier, S., et al. 2015, Icarus 250, 95}
\bibitem[Vuitton et al.(2007)]{vuitton2007}{Vuitton, V., Yelle, R.~V., McEwan, M.~J.\ 2007, Icarus 191, 722}
\bibitem[Wolkenberg et al.(2018)]{wolkenberg2018}{Wolkenberg, P., Smith, M.D., Formisano, V., Sindoni, G. 2018, Icarus 215, 628}
\bibitem[Zhang et al.(2017)]{zhang2017}{Zhang, Z., et al., 2017. Icarus, 281, 297-321.}
\bibitem[Zhang et al.(2018)]{zhang2018}{Zhang, Z., et al., 2018. Icarus, in press.} 
%\bibitem[]
%\bibitem[
%\bibitem[
%\bibitem[
%See {\footnotesize \url{http://www.somewhere.com/see_there's still_characters_here}}

\end{thebibliography}
% For non-BibTex:

\end{document}